\documentclass[preprint]{aastex}
\RequirePackage{epstopdf}

\bibpunct{(}{)}{,}{a}{}{,}

\begin{document}
\received{}
\revised{}
\accepted{}
\slugcomment{Accepted by ApJ on 14 October 2015}

\shortauthors{Adams et al. }
\shorttitle{SOFIA/FORCAST Observations of S106}

\title{SOFIA/FORCAST Observations of Warm Dust in S106: A Fragmented Environment}

\author{J. D. Adams\altaffilmark{1,2}, T. L. Herter\altaffilmark{2}, J. L. Hora\altaffilmark{3}, N. Schneider\altaffilmark{4},
R. M. Lau\altaffilmark{2}, J. G. Staguhn\altaffilmark{5,6}, R. Simon\altaffilmark{4}, N. Smith\altaffilmark{7}, 
R. D. Gehrz\altaffilmark{8}, L. E. Allen\altaffilmark{9}, S. Bontemps\altaffilmark{10}, S. J. Carey\altaffilmark{11}, 
G. G. Fazio\altaffilmark{3}, R. A. Gutermuth\altaffilmark{12}, A. Guzman Fernandez\altaffilmark{3}, M. Hankins\altaffilmark{2}, 
T. Hill\altaffilmark{13}, E. Keto\altaffilmark{3}, X. P. Koenig\altaffilmark{14}, 
K. E. Kraemer\altaffilmark{15},  S. T. Megeath\altaffilmark{16}, D. R. Mizuno\altaffilmark{15}, 
F. Motte\altaffilmark{17}, P. C. Myers\altaffilmark{3}, H. A. Smith\altaffilmark{3}}

\altaffiltext{1}{Stratospheric Observatory for Infrared Astronomy, Universities Space Research Association, NASA/Armstrong Flight Research Center, 2825 East Avenue P, Palmdale, CA 93550, USA}
\altaffiltext{2}{Department of Astronomy, Cornell University, Space Sciences Bldg., Ithaca, NY, 14853, USA}
\altaffiltext{3}{Harvard-Smithsonian Center for Astrophysics, 60 Garden St., Cambridge, MA 02138, USA}
\altaffiltext{4}{KOSMA, I. Physikalisches Institut, Universit\"{a}t zu K\"{o}ln, Z\"{u}lpicher Str. 77, 50937 K\"{o}ln, Germany}
\altaffiltext{5}{NASA/Goddard Space Flight Center, 8800 Greenbelt Road, Greenbelt, MD 20771, USA}
\altaffiltext{6}{Department of Physics and Astronomy, Johns Hopkins University, 3400 N. Charles Street, Baltimore, MD 21218, USA}
\altaffiltext{7}{Department of Astronomy, University of Arizona, 933 North Cherry Avenue, Tucson, AZ 85721-0065, USA}
\altaffiltext{8}{Minnesota Institute for Astrophysics, University of Minnesota, 116 Church St. SE, Minneapolis, MN 55455, USA}
\altaffiltext{9}{National Optical Astronomy Observatory, 950 North Cherry Avenue, Tucson, AZ 85719, USA}
\altaffiltext{10}{Universit\'{e} Bordeaux, LAB, UMR 5804, CNRS, 33270, Floirac, France}
\altaffiltext{11}{Spitzer Science Center, California Institute of Technology, 1200 East California Boulevard, Pasadena, CA 91125, USA}
\altaffiltext{12}{Department of Astronomy, University of Massachusetts, LGRT-B 619E, 710 North Pleasant Street, Amherst, MA 01003-9305, USA} 
\altaffiltext{13}{Joint ALMA Observatory, 3107 Alonso de Cordova, Vitacura, Santiago, Chile}
\altaffiltext{14}{Department of Astronomy, Yale University, New Haven, CT 06511, USA} 
\altaffiltext{15}{Institute for Scientific Research, Boston College, Chestnut Hill, MA 02467, USA}
\altaffiltext{16}{Department of Physics and Astronomy, University of Toledo, 2801 West Bancroft Street, Toledo, Ohio 43606, USA}
\altaffiltext{17}{Laboratoire AIM Paris Saclay, CEA/Irfu - Universit\'{e} Paris Diderot - CNRS, Centre d'\'{E}tudes de Saclay, 91191 Gif-sur-Yvette, France}

\begin{abstract}
We present mid-IR (19 -- 37 $\mu$m) imaging observations of S106 from SOFIA/FORCAST, complemented with IR observations from
{\it Spitzer}/IRAC (3.6 -- 8.0 $\mu$m), IRTF/MIRLIN (11.3 and 12.5 $\mu$m), 
and {\it Herschel}/PACS (70 and 160 $\mu$m). We use these observations, observations in the literature, and 
radiation transfer modeling to study the heating and composition of the warm ($\sim 100$ K) dust in the region. 
The dust is heated radiatively by the source S106 IR, with little contributions
from grain-electron collisions and Ly-$\alpha$ radiation. 
The dust luminosity is $\gtrsim (9.02 \pm 1.01) \times 10^4~\rm{L}_\odot$, consistent with heating
by a mid- to late-type O star.
We find a temperature gradient ($\sim 75 - 107$ K) in the lobes, 
which is consistent with a dusty equatorial geometry around S106 IR. Furthermore, the SOFIA observations resolve 
several cool ($\sim 65 - 70$ K) lanes and pockets of warmer ($\sim 75 - 90$ K) dust in the ionization shadow, indicating 
that the environment is fragmented. We model the dust mass as a composition of amorphous silicates, amorphous carbon,
big grains, very small grains, and PAHs. We present the relative abundances of each grain component for several locations
in S106.
\end{abstract}


\section{INTRODUCTION}

Dust plays a critical role in the life cycle of stars and the interstellar medium. Stars form from the gravitational collapse of
gas and dust in the interstellar medium, while dust that forms in the ejecta of evolved stars and supernovae
enriches the interstellar medium. Dust from evolved stars and supernovae is later available as star-forming material. In star formation
regions, the dust both shields the cold inner regions of dense clouds, leading to the conditions necessary for collapse, and can 
be subsequently heated and processed by stars as they form. Since this dust is a building block for protostellar envelopes
and extrasolar planetary systems, the composition and processing of the dust by stellar radiation is worthy of study.

S106 \citep{sharpless59} is a well-studied, large ($\sim 3^\prime$), bipolar H {\small II} region \citep{pipher78, israel78, lucas78, tokunaga79, herter82}
excited by a luminous source \citep{sibille75, allen75, gehrz82}, referred to as S106 IR. 
Recent analysis performed by \citet{schneider07} suggests that S106 is part of the Cygnus-X complex and is located at a distance of 
$\sim 1.4$ kpc, a distance that is substantially farther than some earlier distance estimates of $\sim 600$ pc 
\citep{eiroa79, staude82} and placing it closer to certain OB associations that can affect its surrounding molecular cloud 
with radiation and winds. S106 presents an opportuntity for us to study the composition
and heating of dust in the vicinity of a massive, luminous star.

The effective temperature (37,000 -- 40,000 K) of S106 IR has been determined by analysis of the observed
emission line intensities in the ionized region (van den Ancker, Tielens, \& Wesselius \citeyear{vandenancker00}). 
This temperature corresponds to a spectral type of
O6 --O9, depending on the method of calibration applied to model stellar atmospheres that are used for reference
\citep{schaerer96, stasinska97, martins05}.

Previous work also includes the characterization of the molecular cloud and heating of the gas \citep{schneider02, schneider03},
as well as the PAH and warm ($\sim 130$ K) dust components \citep{gehrz82, smith01} throughout the bipolar nebula. 
The dust in the bipolar region is concentrated in several bright, compact sources, some of which lie along the bipolar limb. 
The dust is heated radiatively by S106 IR. In the ionzed region, there is the possibility for further heating from collisions
between grains and electrons, as well as Ly-$\alpha$ radiation \citep{smith01}. However, the relative contribution to the 
total dust heating by the non-stellar heating sources have yet to be examined.

It has been known for decades that material around S106 IR casts a UV shadow seen in high spatial resolution radio 
contiuum images (Bally, Snell \& Predmore \citeyear{bally83}).  The UV shadow bisects the bipolar lobes. Near-IR dark lanes \citep{hodapp91, oasa06}, polycyclic aromatic hydrocarbon (PAH) emission \citep{smith01}, and H $\alpha$ emission \citep{bally98} are all present within this shadow.

Coincident with the dark lanes is emission from a cold dust bar \citep{mezger87} and recently detected warm CO gas
\citep{simon12}. Interpretations of the earlier observations invoked the existence of a large ($\sim 30^{\prime\prime} - 60^{\prime\prime}$), continuous disk around 
S106 IR \citep{bally82, bieging84, mezger87}. However, higher angular resolution observations presented by \citet{barsony89} 
failed to detect a large, continuous disk containing molecular gas and instead revealed molecular gas fragments. A similar result
for the cold dust was confirmed by \citet{richer93}, whose millimeter observations showed that the cold dust was broken up into 
several sources. Recent observations of molecular gas velocity from \citet{schneider02} do not provide evidence for a large
smooth or fragmented disk.

Circumstellar material has been detected at very small distances from S106 IR.
VLA observations presented by \citet{gibb07} provide direct evidence for an ionization wind from circumstellar material 
located very close ($\sim 60$ AU) to S106 IR. Emission from this region is 
elongated in the directions that are perpendicular to the bipolar lobes. However, \citet{simon12} showed that there is also a 
column of warm, dense CO gas which contributes to the extinction of the stellar flux in addition to the possible existence 
of a small disk. CO $\nu=2-0$ bandhead emission, observed and modeled by \citet{murakawa13}, is confined
to a ring at 0.3 -- 4 AU from S106 IR, assuming a stellar mass of 20 M$_\odot$. These CO $\nu=2-0$ observations provide 
compelling evidence for a disk on AU scales.

In this paper, we present new IR observations from state-of-the-art ground-based, airborne,
and space-based facilities, with the aim of characterizing the dust heating and composition. We use radiative transfer 
modeling to explain the dust equilibrium 
temperatures and identify a plausible mineral composition for the dust grains. Finally, we discuss the implications
that our observations and modeling have for the nature of S106 IR, contribution to the dust heating from non-stellar
sources, dust composition in the bipolar region, and the H {\small II} region.

\section{OBSERVATIONS}\label{sec:observations}

\subsection{{\it Spitzer}/IRAC}
S106 was observed with {\it Spitzer}/IRAC \citep{werner04, fazio04} as part of the IRAC Guaranteed Time Observations (GTO) program 
(PID 6, AOR 3657472) and during the Cygnus-X Legacy Survey (PID 40184, AORs 27108352, 27107328, 27108608; Hora et al. \citeyear{hora09}, Kraemer et al. \citeyear{kraemer10}) at wavelengths of 3.6, 4.5, 5.8, and 8.0 $\mu$m. The images were obtained using the 12 s high dynamic range mode, which takes one short (0.6 s) and one long (12 s) frame at each pointing in the map.  Part of the region was also observed in the GLIMPSE360 Exploration Science program during the Spitzer Warm Mission at 3.6 and 4.5 $\mu$m using 12 s frames (PID 61072, AOR 42053120; PI B. Whitney). We used the latest version of the Basic Calibrated Data (BCD) available in the archive\footnote{http://irsa.ipac.caltech.edu/data/SPITZER/docs/spitzerdataarchives/}, which was version 18.7.0 except for AOR 42053120 which was version 18.18.0. The BCD were used rather than the ``corrected'' or cBCD because the pipeline automatic column pulldown correction causes many artifacts in the data near regions of bright extended emission, as is present in S106. The IRAC images were individually processed using the routine imclean\footnote{See http://irsa.ipac.caltech.edu/data/SPITZER/docs/dataanalysistools/tools/contributed/irac/imclean/} which is an IRAF\footnote{IRAF is distributed by the National Optical Astronomical Observatories, operated by the Association of the Universities for Research in Astronomy, Inc., under cooperative agreement with the National Science Foundation.}-based script for removing the bright source artifacts (``pulldown'', ``muxbleed'', and ``banding'') \citep{hor04, pip04} from the BCD images. Saturated regions and the resulting artifacts in the 12 s frames were masked as well, allowing the unsaturated short frame data to be used in those locations. The artifact-corrected BCDs were mosaicked into larger images using the IRACproc package \citep{sch06}. IRACproc is a PDL script based on the Spitzer Science Center's post-BCD processing software MOPEX \citep{mah05} which has been enhanced for better cosmic ray rejection. The final images have plate scales of 0\farcs863 per pixel, with intensity units in MJy/sr.

\subsection{IRTF/MIRLIN}
We include observations taken with the MIRLIN camera \citep{ressler94} at
the 3m NASA Infrared Telescope Facility (IRTF) on June 13, 2002. These
observations include images taken through the CVF at a wavelength of
11.300 $\mu$m and through the N5 filter at a center wavelength of 12.492
$\mu$m. The MIRLIN plate scale at the IRTF is 0\farcs47 per pixel on the
128$\times$128 Si:As BIB array. The observations
were taken in chop/nod mode with cumulative integration times of 360 s and
112.5 s for 11.300 $\mu$m and 12.492 $\mu$m, respectively. The chopper throw was
30{\arcsec} east/west with a similar north/south telescope nod that we varied
slightly from one set of images to the next. After sky subtraction,
the many individual frames were resampled to a smaller pixel
scale (0\farcs12 per pixel) and then shifted and added using the bright central star for
spatial registration. The observations were flux-calibrated using
observations of $\beta$ Peg taken immediately after S106, adopting the
zero-magnitude fluxes in the MIRLIN handbook.

\subsection{SOFIA/FORCAST}
We observed S106 with the FORCAST instrument \citep{adams12a, herter12} on SOFIA \citep{young12} 
at the wavelengths 19.7, 25.3, 31.5, and 37.1 $\mu$m, using FORCAST guaranteed observing time during 
Cycle 1 (Flight 108 on June 21, 2013) and Cycle 2 (Flight 170 on May 8, 2014).  FORCAST is a dual channel imager
and spectrometer that utilizes a $256 \times 256$ pixel format Si:As detector for 5 -- 26 $\mu$m and a $256 \times 256$
pixel format Si:Sb detector for 26 -- 40 $\mu$m, with a rectified plate scale of 0\farcs768 per pixel in each camera.
The wavelengths 19.7, 25.3, and 31.5 $\mu$m were observed using the dichroic 
beamsplitter, while 37.1 $\mu$m was observed directly. The observations were performed with off-source chopping and nodding 
(C2NC2 mode) and dithering, whereby the off-source fields were chosen to be ones free of detectable emission in 
MSX \citep{price01} 21 $\mu$m images. Typical dwell times in each nod beam/dither position were approximately 30 -- 90 sec,
including chopping inefficiencies.

The raw data were reduced and calibrated using the pipeline described in \citet{herter13}. This pipeline
performs chop and nod subtraction and applies corrections for droop, detector non-linearity, multiplexer crosstalk, and optical
distortion. Calibration factors were derived from the average photometric instrument response to standard stars observed 
during each corresponding observing campaign (a series of four or more flights). The dither positions were aligned and averaged, producing 
images with effective on-source exposure times of 480, 371, 375, and 195 sec for 19.7, 25.3, 31.5, and 37.1 $\mu$m, respectively. 
The rectified plate scale for all the FORCAST images was
$0.768^{\prime\prime}$ per pixel. The effective RMS noise levels were approximately 8, 10, 10, and 13 mJy/pixel for
19.7, 25.3, 31.5, and 37.1 $\mu$m, respectively. We estimate the calibration error due to flat fielding to be $\sim 10\%$.

We performed image deconvolution on the FORCAST images using several iterations of a maximum likelihood algorithm 
\citep{richardson72, lucy74} to produce images with an effective spatial resolution of $\sim 2.4^{\prime\prime}$. 
The application of this algorithm requires a point spread function (PSF) to be used as input. 
We chose to use a synthetic PSF since there were no point sources in the field that
were sufficiently bright to use as a reliable reference PSF. The synthetic PSF was a convolution
of an Airy function for telescope diffraction and a two-dimensional Gaussian for pointing instability that was scaled so that
the FWHMs of the final PSF matched those of the point sources in the images that were to be deconvolved
\citep{debuizer12}.

Finally, we derived astrometric solutions (WCS) for the images 
using the location of three point sources in the field and a tangent plane projection. This method produces astrometry that
is accurate to within $\sim 0.4^{\prime\prime}$ \citep{adams12b}. The astrometrically calibrated frames were then aligned
with and sampled to that of the IRAC 3.6 $\mu$m image (0\farcs863 per pixel), whose astrometric solution is determined from the position of the field stars in the 2MASS catalog.

\subsection{{\it Herschel}/PACS}
In this paper, we use {\it Herschel}/PACS \citep{poglitsch10} 70 and 160 $\mu$m observations
from the {\it Herschel} Open-Time program OT2\_jhora\_2 (obsIDs
1342257386 and 1342257387). The data were taken on December 18, 2012,
simultaneously with SPIRE \citep{griffin10} and PACS,
using a scanning speed of 60$''$/s with one repetition in the
nominal and orthogonal observing directions.  The individual scan
directions of the PACS data were first reduced from the raw data of
level-0 to level-1 using the standard pipeline in HIPE 10.2751. This
includes flat-field and non-linearity corrections. The next step
(from level-1 to level-2) was performed within Scanamorphos v.21
\citep{roussel13}. The processing consists of subtracting long and short
timescale drifts and masking glitches and brightness
discontinuities. The two scan directions were merged and projected
onto a spatial grid of 2$''$/pixel and 3$''$/pixel for 70 $\mu$m and
160 $\mu$m, respectively. The angular resolution of the data is
6$''\times$12$''$ for 70$\mu$m and 12$''\times$16$''$ for 160 $\mu$m,
respectively (PACS Manual v.2.5.1). The absolute calibration uncertainties 
for the integrated source flux densities are estimated to be $\sim 10\%$ for 70 
$\mu$m and $\sim 20\%$ for 160 $\mu$m. 

\section{OBSERVATIONAL RESULTS}\label{sec:results}

\subsection{Morphology}

\begin{figure}
\plotone{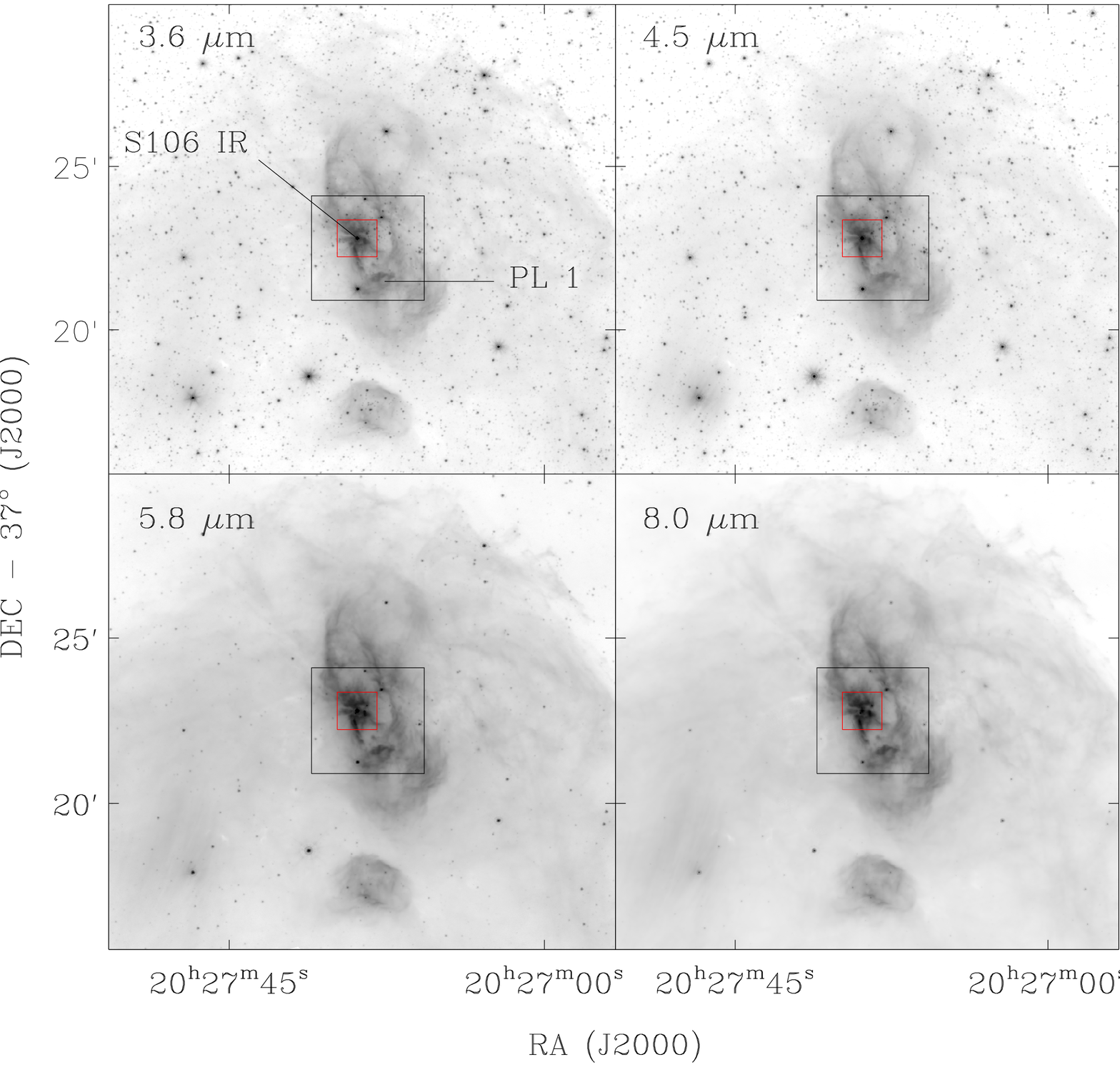}
\caption{Combined long (12 sec) and short (0.6 sec) exposure IRAC images of S106 at 3.6, 4.5, 5.8, and 8.0 $\mu$m.  
The displayed image intensities are stretched logarithmically. The locations of S106 IR and the clump PL 1 are indicated in the upper left panel.
The black boxed region denotes the size of the FORCAST field-of-view 
(without showing the observed SOFIA field rotations), while the red boxed region shows the MIRLIN field-of-view.
These images trace ionization lines,
PAH emission, and thermal continuum emission \citep{vandenancker00}, as well as young stellar objects and photospheres.
\label{fig:spitzerim}}
\end{figure}

Fig. \ref{fig:spitzerim} shows the {\it Spitzer}/IRAC images. Based on the spectrum of S106 presented by van den Ancker et al. (\citeyear{vandenancker00}), these images trace dust continuum emission, recombination line emission, and PAH emission.
In addition, the images contain stars with primarily photospheric emission and young stellar objects that exhibit infrared
excess emission. The thermal emission is bipolar. A dark lane is seen in the region where a similar feature is seen in the near-IR 
\citep{oasa06}. There is also emission from a dusty clump (PL 1) to the southwest of the S106 IR as indicated in Fig. \ref{fig:spitzerim}.

\begin{figure}
\plotone{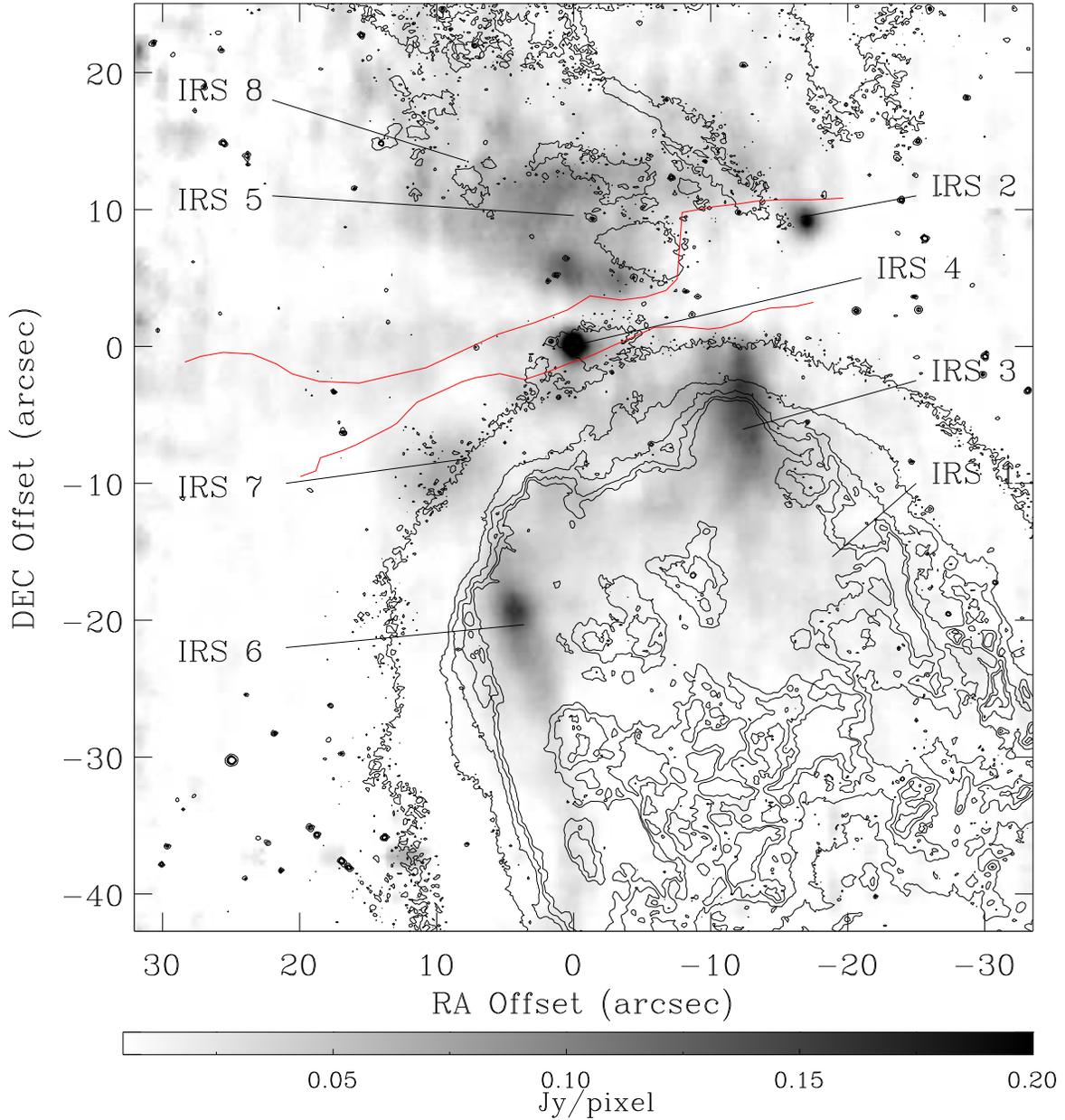}
\caption{IRTF/MIRLIN image of S106 at 11.3 $\mu$m. The image is centered on the coordinates $\rm{RA}=20^{\rm{h}}27^{\rm{m}}26.74^{\rm{s}}\rm{,}~\rm{DEC}=+37^\circ22^\prime48.7^{\prime\prime}$ (J2000), near
S106 IR. The emission in this images is primarily from PAHs. The locations of the IRS sources from \citet{gehrz82} are indicated.
The black contours represent H-$\alpha$ emission \citep{bally98}. The region between the red lines is devoid of 5 GHz emission \citep{bally83}.
\label{fig:mirlin113}}
\end{figure}

\begin{figure}
\plotone{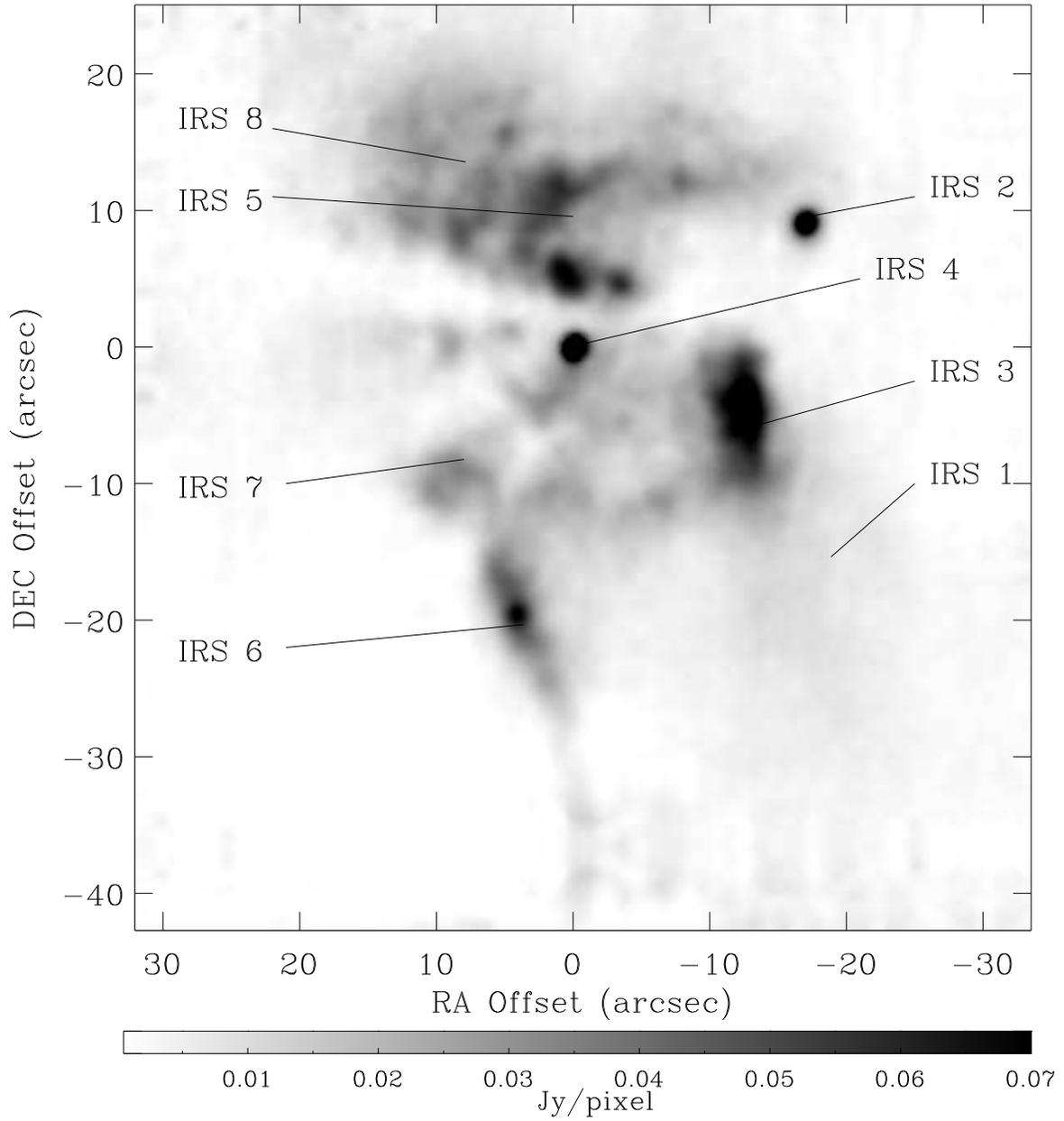}
\caption{Same as Fig. \ref{fig:mirlin113}, but for 12.492 $\mu$m, which traces the dust thermal continuum.
\label{fig:mirlin12}}
\end{figure}

The MIRLIN 11.3 $\mu$m and 12.492 $\mu$m images are shown in Figs. \ref{fig:mirlin113} and \ref{fig:mirlin12}, respectively.
Although the 11.3 $\mu$m image is not continuum subtracted, the emission is dominated by PAHs, while emission at 12.492 $\mu$m traces 
that of the dust thermal continuum. Emission can be seen 
from the bright, compact sources (IRS 1 -- 8) listed in \citet{gehrz82}, but also from the lobes. The dark lane seen in near-IR images is
also dark at both these wavelengths. In Fig. \ref{fig:mirlin113}, we show the locations of the ionized region using H-$\alpha$ contours
\citep{bally98} and the UV shadow as the region devoid of 5 GHz free-free emission \citep{bally83}. The position PL 1 lies outside the MIRLIN field.

\begin{figure}
\plotone{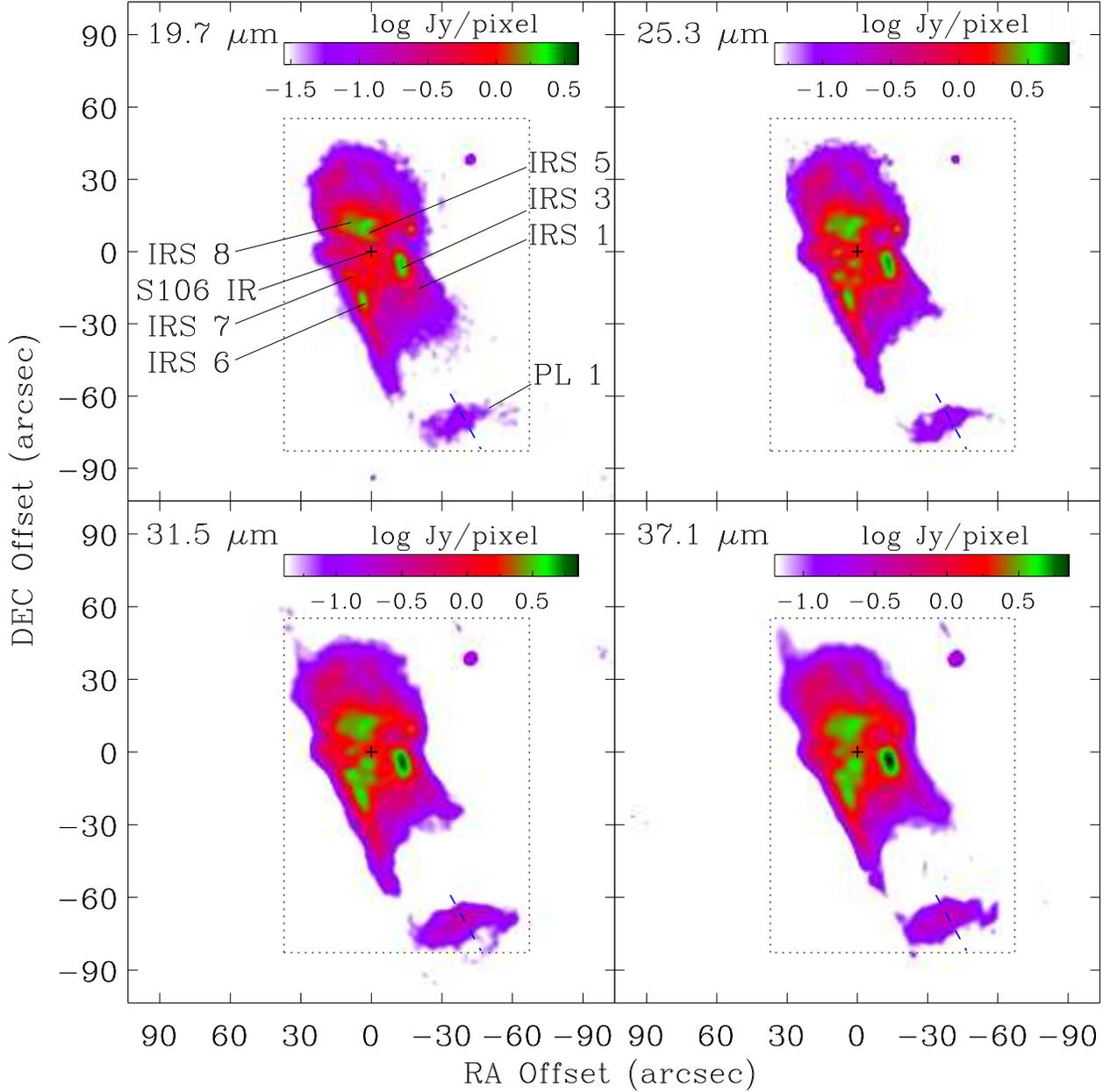}
\caption{Deconvolved SOFIA/FORCAST images of S106 at 19.7, 25.3, 31.5, and 37.1 $\mu$m. 
The image is centered on the coordinates $\rm{RA}=20^{\rm{h}}27^{\rm{m}}26.74^{\rm{s}}\rm{,}~\rm{DEC}=+37^\circ22^\prime48.7^{\prime\prime}$ (J2000), near
S106 IR, with the North direction up and East direction to the left. The deconvolved beam size is $2.4^{\prime\prime}$. 
Edge-of-frame artifacts outside the emission regions have been removed for display purposes. 
Extended sources (IRS 1, 3, 5, 6, 7, 8) detected by \citet{gehrz82} at $\sim 10~\mu$m and PL 1 are identified in the upper left panel. 
The dotted box represents the aperture that was used to compute the total flux (\S\ref{sec:totallum} and Table \ref{tab:totalflux}).
The blue dashed line indicates the location of the profile shown in Fig. \ref{fig:pillarcuts}.
\label{fig:sofiaim}}
\end{figure}

The deconvolved  SOFIA/FORCAST images are shown in Fig. \ref{fig:sofiaim}. Emission was detected from both lobes and from
several bright, compact sources. Several of these sources (IRS 1, IRS 3, IRS 5, IRS 6, IRS 7, IRS 8) were detected in ground-based 
images at $\sim 10~\mu$m and $\sim 18~\mu$m by \citet{gehrz82} and \citet{smith01} and are indicated in Fig. \ref{fig:sofiaim}. 
We did not detect the photosphere of
S106 IR (IRS 4) at 19 -- 37 $\mu$m. The dusty clump PL 1 was detected at all wavelengths by FORCAST.

\begin{figure}
\plotone{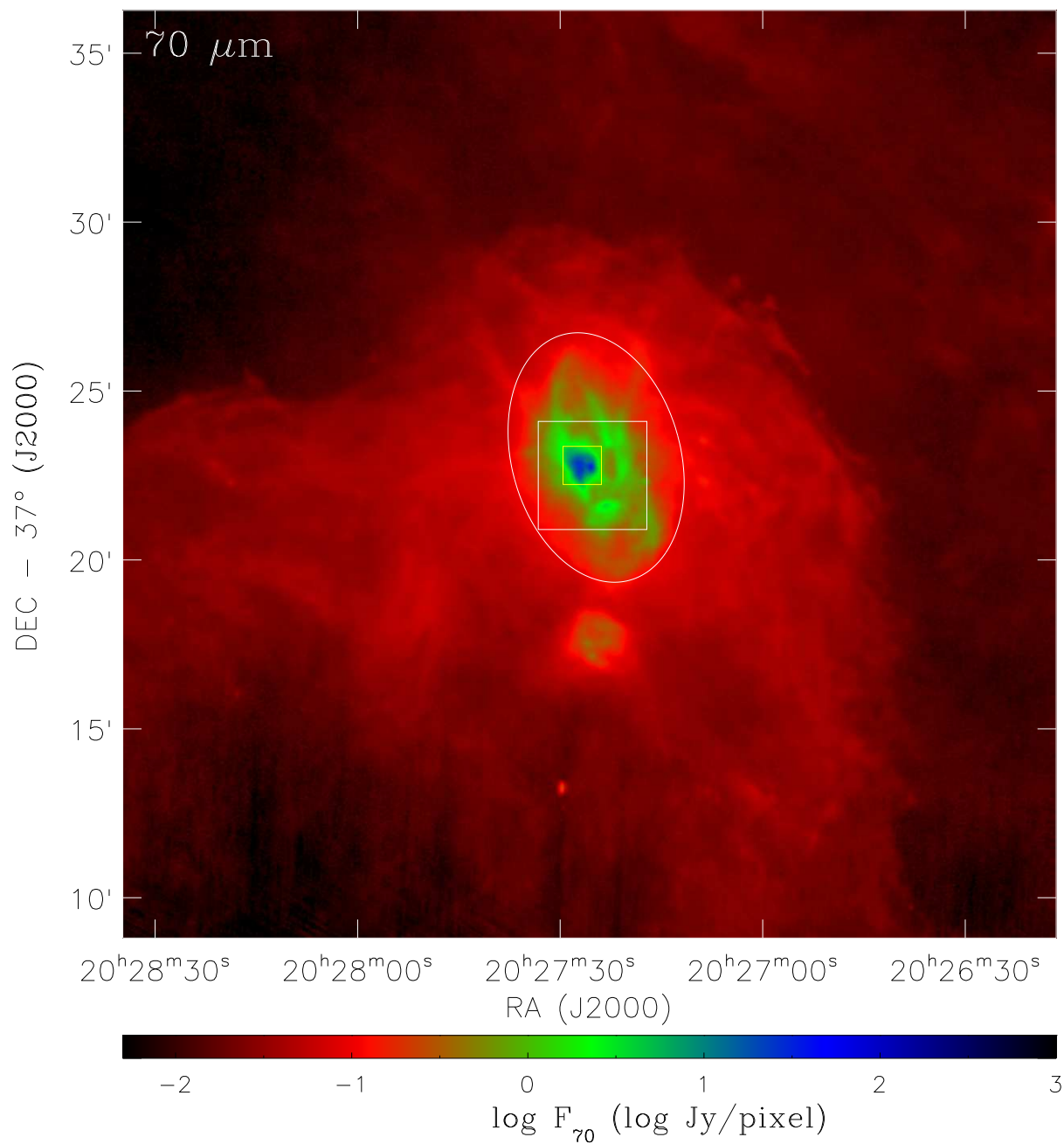}
\caption{Herschel/PACS image of S106 at 70 ${\bf \mu}${\bf m} from the Cygnus-X Open-Time program (OT2\_jhora\_2). 
The white box represents the FORCAST field-of-view while the region outlined in yellow represents
the MIRLIN field-of-view. The white ellipse represents the aperture that was used to compute the total flux (\S\ref{sec:totallum} and Table \ref{tab:totalflux}).
\label{fig:herim70}}
\end{figure}

\begin{figure}
\plotone{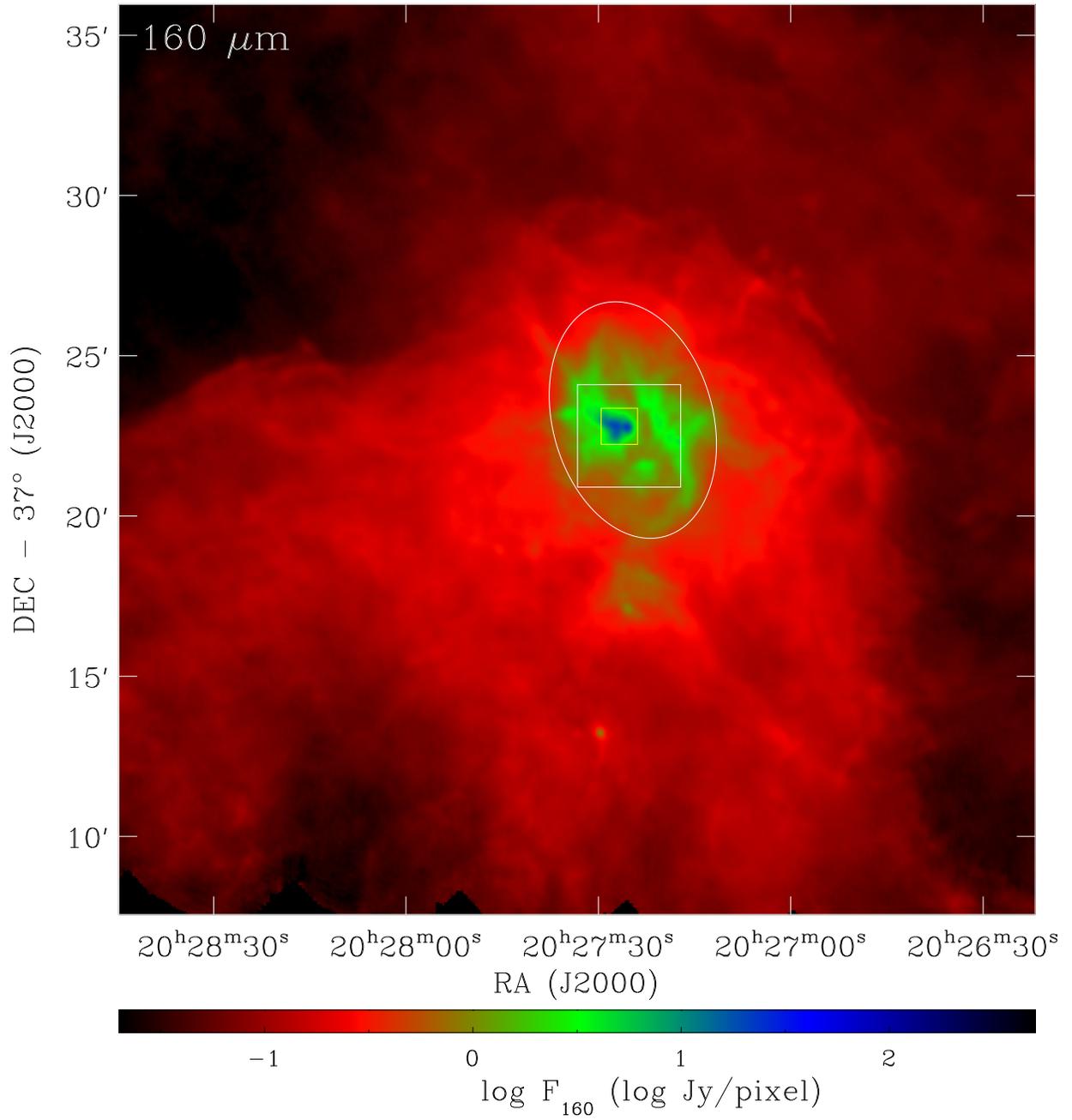}
\caption{Same as Fig. \ref{fig:herim70}, but for 160 $\mu$m. 
\label{fig:herim160}}
\end{figure}

The {\it Herschel}/PACS images are shown in Figs. \ref{fig:herim70} and \ref{fig:herim160}. At 70 $\mu$m, the morphology follows 
that of the warm
dust continuum that is seen at 19 -- 37 $\mu$m.
At 160 $\mu$m, the morphology changes dramatically due to detection of cold dust in the molecular cloud, and resembles
the submillimeter continuum emission \citep{simon12}. A cavity is seen surrounding the western region around the lobes, 
which extends southward along the eastern and western sides of PL 1.

\begin{figure}
\plotone{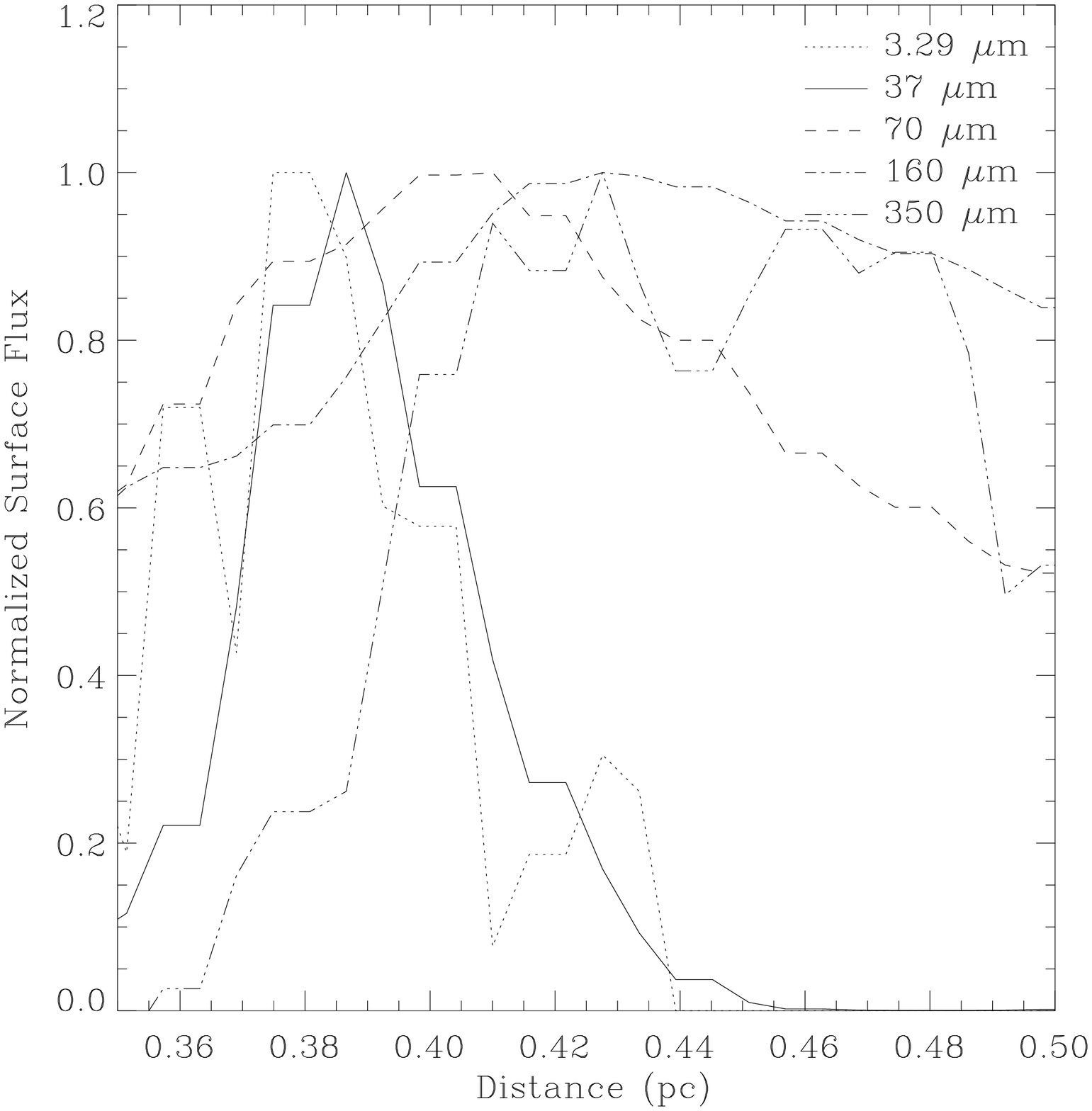}
\caption{Radial profile (as indicated in Fig. \ref{fig:sofiaim}) through the location PL 1 at 3.29 \citep{smith01}, 37, 70, 160, and 350 $\mu$m \citep{simon12}. 
The distance from S106 IR is the projected distance, assuming a distance of 1.4 kpc
to the region \citep{schneider07}. The peak of the PAH emission lies at the inner edge of the clump and is slightly displaced from the
peak 37 $\mu$m emission, while the peak emission shifts to deeper locations in the clump with increasing wavelength. 
\label{fig:pillarcuts}}
\end{figure}

The peak emission at the illuminated edge of PL 1 shifts deeper into the molecular cloud at 160 $\mu$m when compared with
emission at 37 and 70 $\mu$m (Fig. \ref{fig:pillarcuts}), suggesting that the clump is self-absorbing. The amount of this
shift is nearly 0.05 pc in projection. In addition, the peak of the PAH emission \citep{smith01} lies at the inner edge of the clump and
starts to decrease where the 37 $\mu$m emission from larger grains peaks.

\subsection{Dust luminosity}\label{sec:totallum}

In Table \ref{tab:totalflux}, we report the total detected dust continuum flux densities at 12 -- 160 $\mu$m. We
include the flux density at 350 $\mu$m \citep{simon12}.
The aperture was chosen for each wavelength according to field-of-view and the extent of the detected dust emission.
We limited the aperture at 70 and 160 $\mu$m in order to minimize the chance of the inclusion of dust that may be heated by external stars.
We examined the 70/160 $\mu$m color temperature in this aperture and did not find a gradient that would indicate external heating. 
The apertures for the FORCAST and Herschel images are shown in Figs. \ref{fig:sofiaim} -- \ref{fig:herim160}.
The flux densities of the point sources IRS 2 and IRS 4 have been subtracted (where applicable). From these flux densities, 
we derive a lower limit to the total dust
luminosity of $L_{dust} \gtrsim (9.02 \pm 1.01) \times 10^4~\rm{L}_\odot$. We state the luminosity as a lower limit because
the some of the stellar radiation may be escaping through the bipolar geometry and there may be dust emission outside the
70 and 160 $\mu$m apertures.

\subsection{19/37 $\mu$m color temperature}

\begin{figure}
\plottwo{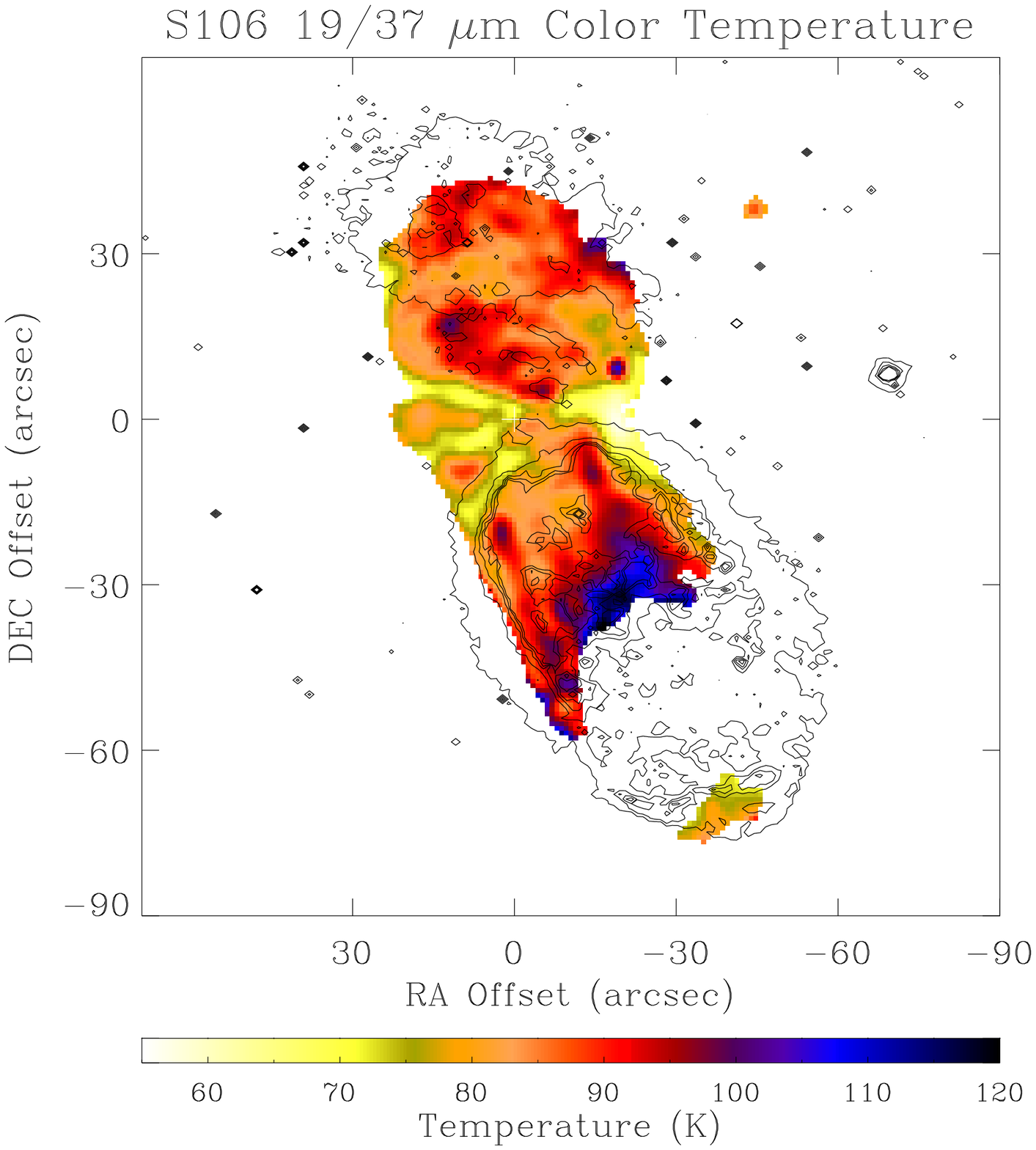}{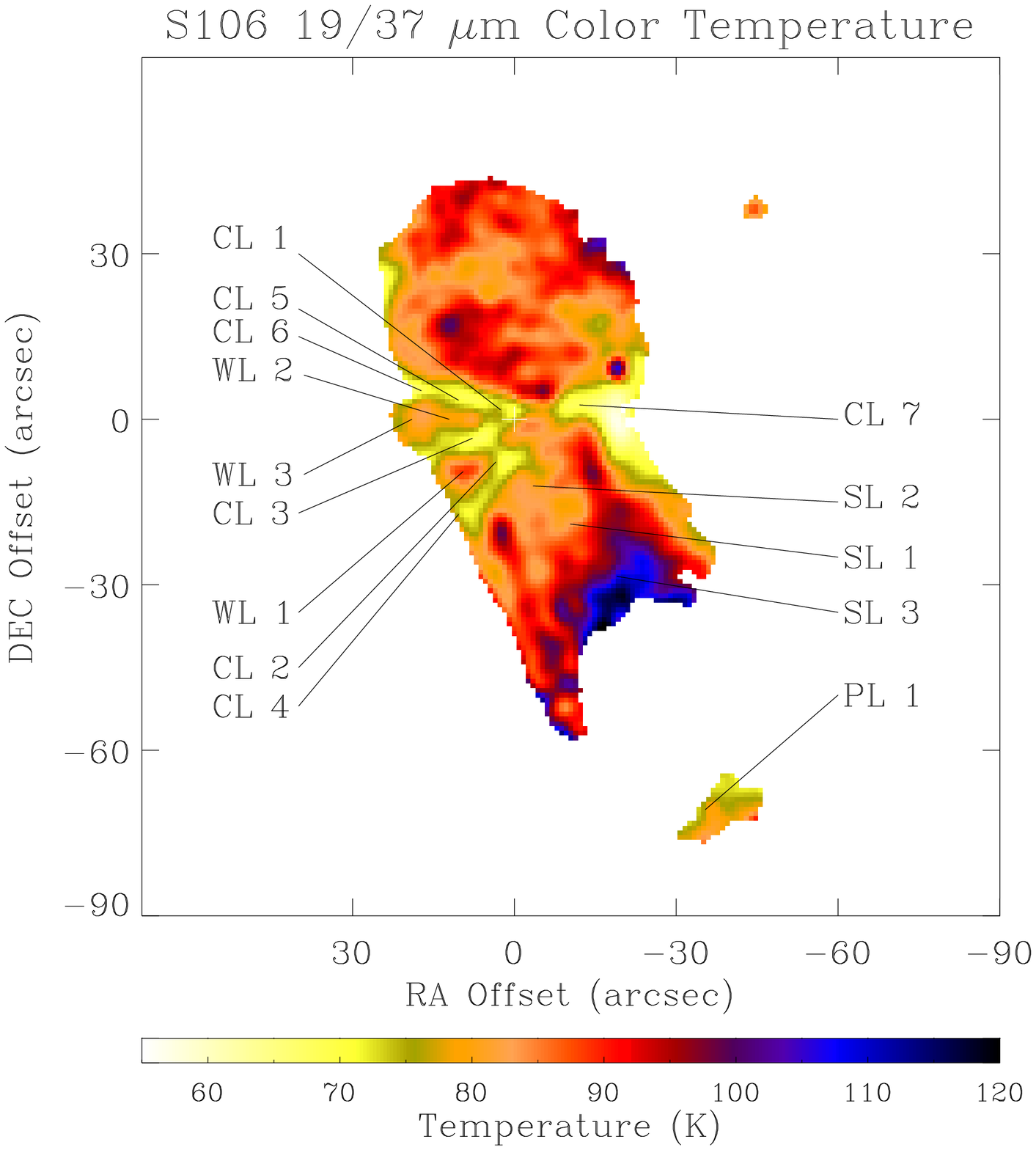}
\caption{Color-temperature map derived from the FORCAST 19.7 and 37.1 flux maps, assuming a dust emissivity law that is proportional to 
$\lambda^{-1.8}$ \citep{abergel11}. The coordinates of the map are centered near S106 IR and the coordinates
are given in the caption to Fig. \ref{fig:sofiaim}. {\it Left}: Color-temperature map, overlaid with contours of H-$\alpha$ emission \citep{bally98}.
The UV shadow cone is resolved into cool lanes separated by compact regions of warmer dust. 
The hottest dust is found in the lobes. {\it Right}: The color temperature map with representative dust modeling positions (see \S\ref{sec:modeling}) indicated.
\label{fig:ctemp}}
\vspace{5.0cm}
\end{figure}

A modified blackbody color temperature map derived from the 19 and 37 $\mu$m flux ratios is shown in Fig. \ref{fig:ctemp} with
H-$\alpha$ emission \citep{bally98} overlaid with contours. 
For this map, the emissivity of the dust grains was assumed to be proportional to $\lambda^{-1.8}$ \citep{abergel11}. Overall, the 
dust temperatures are lower 
($\sim 70 - 90$ K) in the equatorial region and tend to increase out into the lobes ($\sim 107$ K). 

We resolved several cool lanes in the S106 IR UV shadow, with temperatures of $\sim 70$ K, and three compact regions in the UV shadow
with temperatures of $\sim 90$ K. One of these warmer regions was identified as a 10 $\mu$m compact source (IRS 7) by \citet{gehrz82}.

Fig. \ref{fig:ctemp} also depicts the locations used for dust modeling (\S\ref{sec:modeling}). The positions and fluxes at these locations are listed
in Table \ref{tab:fluxes}.

\subsection{37 $\mu$m optical depth}

\begin{figure}
\plotone{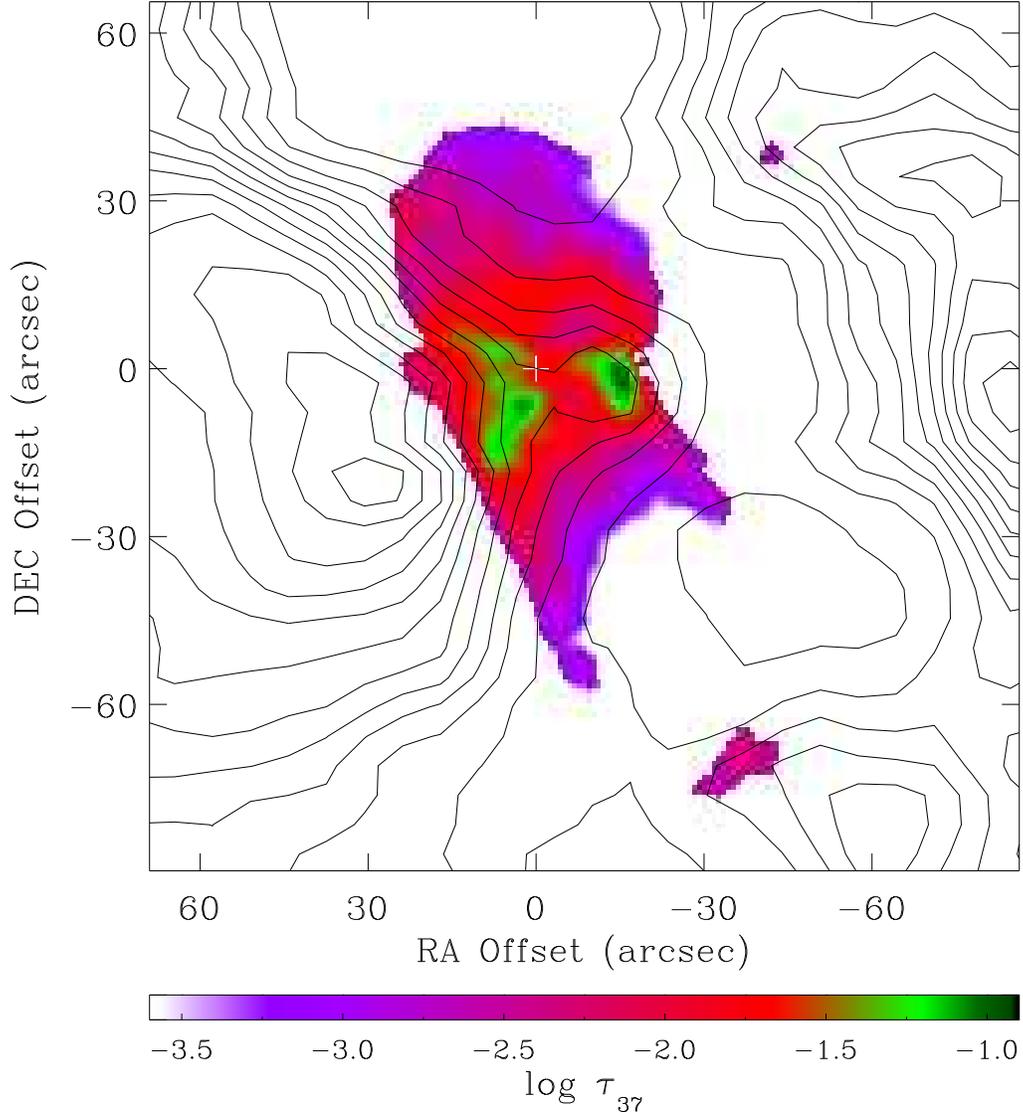}
\caption{Optical depth $\tau$ at 37 microns along the line of sight, derived from the temperatures displayed 
in Fig. \ref{fig:ctemp}. The coordinates of the map are centered near S106 IR and the coordinates
are given in the caption to Fig. \ref{fig:sofiaim}. The countours represent velocity-integrated $^{13}$CO (2-1) emission from IRAM observations \citep{schneider02}. The CO data have a spatial resolution of 
$\sim 11^{\prime\prime}$ and the contour levels are spaced through the range 20 -- 80 Jy/beam. 
\label{fig:tau}}
\end{figure}

Using the temperature map shown in Fig. \ref{fig:ctemp} and the 37 $\mu$m image, we computed the optical depth $\tau$ along the line 
of sight. We show this optical depth map in Fig. \ref{fig:tau}. The highest column densities ($0.08 \lesssim \tau \lesssim 0.12$)
are located near S106 IR, in the UV shadow,
with the lobes becoming increasingly optically thin ($\tau \lesssim 0.08$) at larger distances in the lobes. 
The FORCAST images have low signal-to-noise ratios between IRS 6 and PL 1, indicating a low dust column density or absence of dust in
this area.
 
\section{DUST MODELING}\label{sec:modeling}

We used radiation transfer to model the spectral energy distribution (SED) of the dust at the locations specified in Fig. \ref{fig:ctemp}.
Non-stellar components to the dust heating are addressed in \S\ref{sec:discussion}. We used the DustEM
software \citep{compiegne11} to specify the dust model population and compute its IR emission in a stellar
radiation field.

\subsection{Radiation field}
At a location within the nebula, the incident radiation field from the star is $(R_*/D)^2 F_{\lambda,*} e^{-\tau_\lambda}$, where
$R_*$ is the radius of the star, $D$ is the distance from the stellar surface to the dust, $F_{\lambda,*}$ is the flux density of
the star at the surface of the star, and $\tau_\lambda$ is the optical depth of the dust between the star and the location. 
The distance from the star to the dust was set to the projected distance ($D_{proj}$)
between them, scaled to a distance of 1.4 kpc \citep{schneider07}. 
We adopt $\tau_\lambda = \tau_{UV} (121.5~\rm{nm} / \lambda)^\beta$, where
$\beta = 1.85$ \citep{landini84} and $\tau_{UV}$ is a free parameter. For the stellar flux density, we used a stellar 
atmosphere with a spectral type of O7V and an effective temperature of 37,000 K from \citet{castelli04}.

\subsection{Grain types and size distributions}
For the dust component, we consider contributions to the emission from amorphous silicates \citep{draine07}, amorphous carbon,
neutral PAHs, and ionized PAHs \citep{compiegne10}. In cases of relatively low extinction, we consider emission from both very 
small ($1.2 \times 10^{-7} \le a \le 1.5 \times 10^{-6}$ cm, where $a$ is the grain radius) grains (VSGs), which
can be transiently heated in a UV radiation field \citep{ryter87}, and big ($1.5 \times 10^{-6} \le a \le 1.1 \times 10^{-5}$ cm) 
grains (BGs), which are thought to be
composed of silicates \citep{desert90}. The grain size distribution for both PAH species is modeled as
\begin{equation}
\frac{dn}{d{\rm log}a} \propto \frac{e^{-{\rm log}(\frac{a}{a_0})^2}}{\sigma}
\end{equation}
where $3.1 \times 10^{-8}~{\rm cm} \le a \le 1.2 \times 10^{-7}~{\rm cm}$, $a_0 = 3.1 \times 10^{-8}~{\rm cm}$,
and $\sigma = 0.4$. The grain size distributions for the amorphous silicates, amorphous carbon, VSGs, and BGs were
modelled as $dn/da \propto a^\alpha$ with $\alpha$ as a free parameter for each species. For the amorphous grains,
$3.1 \times 10^{-8}~{\rm cm} \le a \le 2.0 \times 10^{-4}~{\rm cm}$, for VSGs, 
$1.2 \times 10^{-7}~{\rm cm} \le a \le 1.5 \times 10^{-6}~{\rm cm}$, and for BGs, 
$1.5 \times 10^{-6}~{\rm cm} \le a \le 1.1 \times 10^{-5}~{\rm cm}$.

\subsection{SED modeling}

\begin{figure}
\vspace{-18.5cm}
\plotfiddle{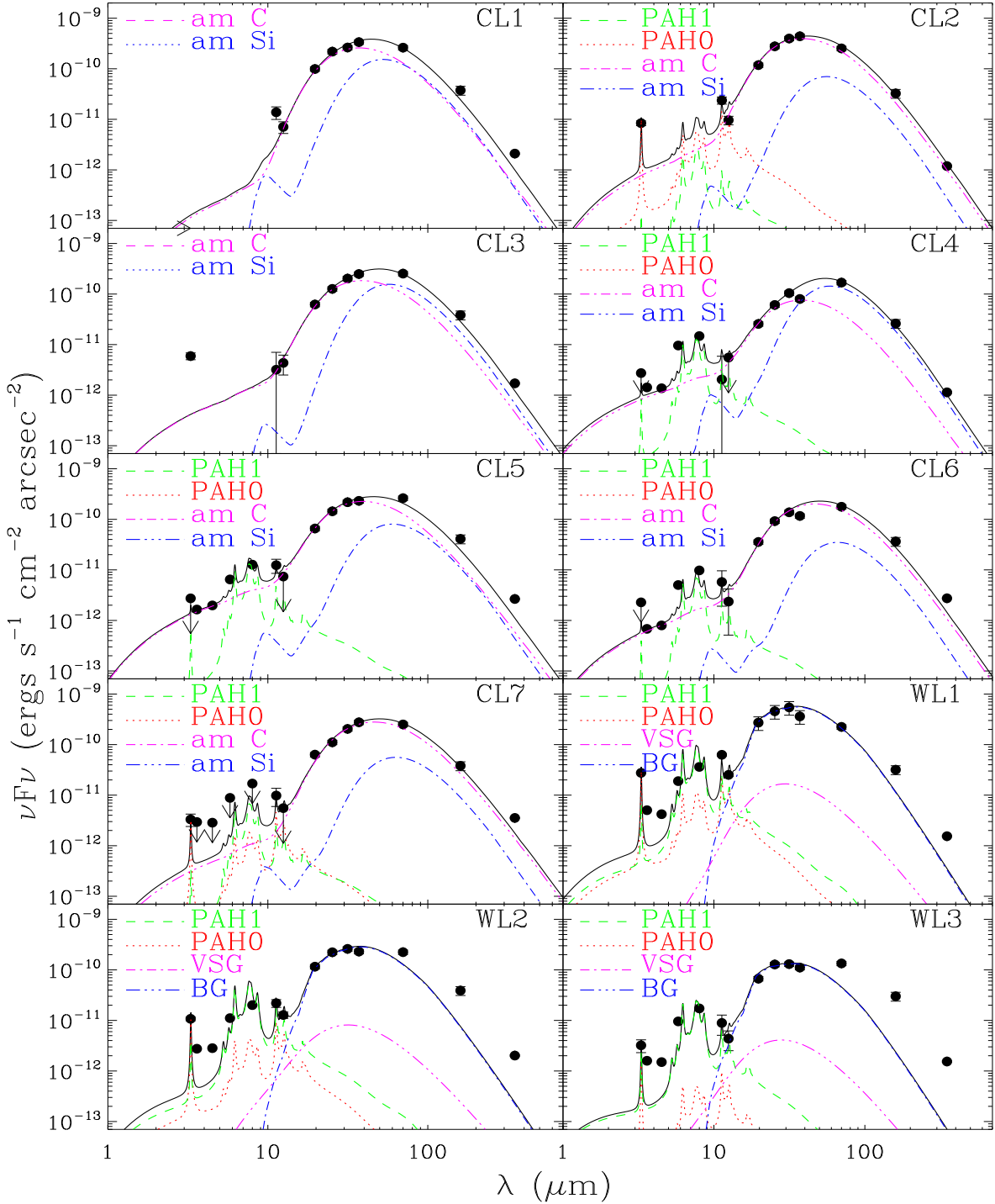}{7.0in}{0.0}{450.}{550.}{0}{0}
\caption{SEDs and model results for locations in cool lanes (CL 1--7) and warm lanes (WL 1--3). These
locations are depicted in Fig. \ref{fig:ctemp}. The total model SEDs are shown as solid black lines.
For CL 1--7, the dust composition is modeled as a mixture of amorphous 
carbon (am C, triple-dot-dashed magenta lines) and amorphous silicates (am Si, dot-dashed blue lines),
which in some cases contain neutral PAHs (PAH0, dotted red lines) and/or ionized PAHs (PAH1, dashed green lines).
For WL 1--3, the dust grain types are changed to VSGs (triple dot-dashed magenta lines), and BGs (dot-dashed blue lines),
and include PAHs as well. For the WL positions, the IRAC 3.6, 4.5, and 8.0 $\mu$m bands will contain emission from ionization 
lines (see text), and were not used to constrain the model components.
\label{fig:models1}}
\end{figure}

\begin{figure}
\vspace{-18.5cm}
\plotfiddle{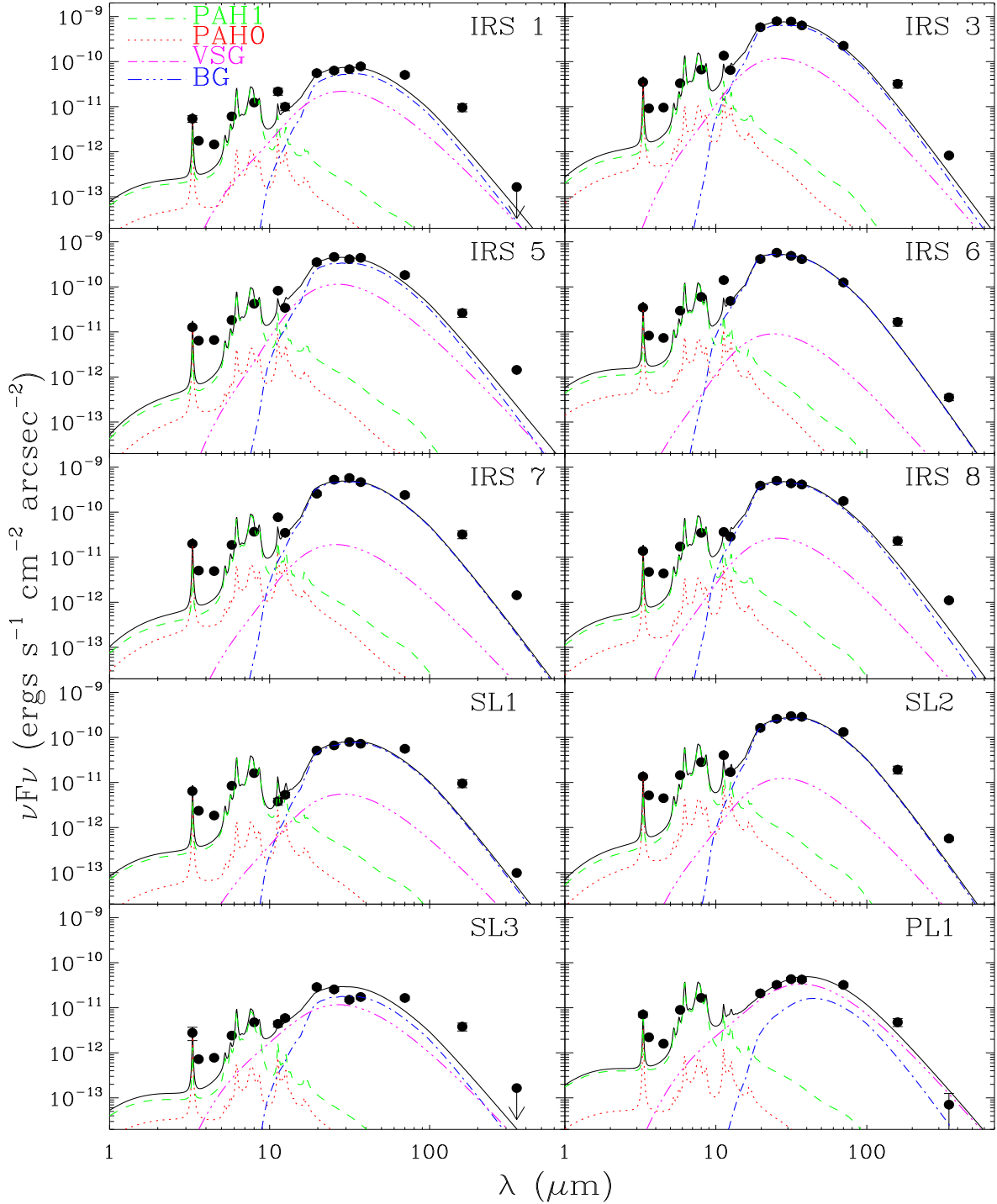}{7.0in}{0.0}{450.}{550.}{0}{0}
\caption{Same as Fig. \ref{fig:models1} for extended IRS sources \citep{gehrz82}, southern lobe
locations SL 1--3, and southwestern clump location PL 1 (Fig. \ref{fig:ctemp}). The model grain types for all these positions are
the same as those used for the WL positions. As mentioned in the Fig. \ref{fig:models1} caption, the IRAC 3.6, 4.5, and 
8.0 $\mu$m bands contains ionization lines (see text) for these positions, and were not used to constrain the model components
at these positions.
\label{fig:models2}}
\end{figure}

In Fig. \ref{fig:ctemp}, we specify locations in the nebula for which we show the SED and perform dust SED modeling.
These locations are cool lane positions CL 1--7, warm lane positions WL 1--3, southern lobe positions
SL 1--3, southwestern clump position PL 1, and the positions of the bright, compact sources IRS 1, IRS 3, IRS 5, IRS 6, IRS 7, 
and IRS 8 identified in \citet{gehrz82}. In Figs. \ref{fig:models1} and \ref{fig:models2}, we show the SEDs for these positions, 
including 3.29 $\mu$m IRTF/NSFCAM data from \citet{smith01} and 350 $\mu$m CSO/SHARC-II data from \citet{simon12}.

For CL 1--7, the dust composition is assumed to be dominated by a mixture of amorphous silicates, 
amorphous carbon, and PAHs (Fig. \ref{fig:models1}). 
Positions CL 1 -- 3 contain contamination from the S106 IR photosphere at 3.29 -- 8.0 $\mu$m, and those data are not shown
in the SEDs. For CL 1, some PAH emission may be
present at 11.3 $\mu$m, but these components are not modeled as the ionization fraction is unconstrained.
For CL 2, we model the PAH components and note that the ionized PAH component is an upper limit based on
emission at 3.29 $\mu$m and 11.3 $\mu$m. At CL 3, there is emission at 3.29 $\mu$m but not at 11.3 $\mu$m. 
In this case, the PAH components are not modeled as the ionization fraction is unconstrained.
For CL 4, we show 3$\sigma$ upper limits at 3.29 and 12.492 $\mu$m. At this position, the IRAC 3.6 and 4.5 $\mu$m points help constrain the 
size distribution for amorphous carbon. Emission at 11.3 $\mu$m is used to place an upper limit on the PAH emission, which underpredicts
the emission at 5.8 $\mu$m. There is a possibility that the IRAC photometry at this position is affected by emission from IRS 6.
At CL 5, the 3$\sigma$ upper limit for 3.29 $\mu$m and 4$\sigma$ upper limit for 12.492 $\mu$m are shown. Again, the IRAC 
fluxes at 3.6 and 4.5 $\mu$m help constrain the size distribution for amorphous carbon.
For CL 6,  a 3$\sigma$ upper limit at 3.29 $\mu$m is shown.
At position CL 7, the IRAC data points are contaminated by emission from IRS3 and are shown as upper limits. A 3$\sigma$ upper limit is shown for 
12.492 $\mu$m.

The modeling parameters and results for the CL positions are given in Table \ref{tab:models1}. 
The modeling results yield values of $\tau_{UV}$ in the range 8 -- 40, with the highest extinction at position CL 1.
We note that $\tau_{UV}$ does {\it not} necessarily increase with increasing distance from S106 IR. The 
amorphous silicates contribute $\sim 50 - 94\%$ to the dust mass, with the remainder contributed by amorphous
carbon ($\sim 6 - 50\%$) and, where detected, PAHs ($< 1\%$).

For WL 1--3, a mixture of PAHs, VSGs, and BGs was modeled (Fig. \ref{fig:models1}). 
The ionized PAH model component was averaged over the IRAC instrument response for the 5.8 $\mu$m bandpass 
and then normalized to
match the emission at 5.8 $\mu$m, which is dominated by the 6.3 $\mu$m PAH feature and contains little recombination
line emission (van den Ancker et al. \citeyear{vandenancker00}). The IRAC 8.0 $\mu$m band contains [Ar II] and [Ar III] 
lines at 6.99 and 8.99 $\mu$m, respectively, and was not used to constrain the model components.
Also, since the IRAC 3.6 and 4.5 $\mu$m bands conain hydrogen recombination lines, such as Pf $\delta$ (3.2970 $\mu$m), 
Pf $\gamma$ (3.7406 $\mu$m), Br $\alpha$ (4.0523 $\mu$m), and Pf $\beta$ (4.6539 $\mu$m) (van den Ancker et al. \citeyear{vandenancker00}), they
were not used for model fitting.
Residual emission at 3.29 $\mu$m above the emission from ionized PAHs was used to normalize the neutral PAH component.
We attribute excess emission at 70 -- 350 $\mu$m to contamination from IRS 6 and emission from cooler
dust along the line of sight that can be seen at both 160 $\mu$m (Fig. \ref{fig:herim70}) and 350 $\mu$m \citep{simon12}.
From Table \ref{tab:models2}, values of $\tau_{UV}$ for the warm lanes are in the range  2--6. The dust mass is dominated
by BGs ($\sim 91 - 98\%$), with contributions from VSGs ($\lesssim 1\%$) and PAHs ($\lesssim 2\%$). The fraction of PAHs
that are ionized spans the range 82 -- 97\%.

SEDs and models for IRS sources, SL 1--3, and PL 1 are shown in Fig. \ref{fig:models2}. We reiterate that PL 1 lies outside
the MIRLIN field. The modeling approach for
these sources was similar to that for WL 1--3, but the results are slightly different (Table \ref{tab:models2}). In
these cases, values for $\tau_{UV}$ are in the range $\sim 0 - 5$, with the lowest values of extinction found
at the southernmost positions, SL 3 and PL 1. Again, the model composition is dominated by BGs ($\sim 50 - 98\%$),
with contributions from VSGs ($\sim 1 - 43\%$) and PAHs ($\sim 2 - 7\%$). The ionized PAH fractions are 88 -- 97\%,
with the higher ionization fraction values correlated with the lower values of $\tau_{UV}$.

\section{DISCUSSION}\label{sec:discussion}

\subsection{The nature of S106 IR}

\subsubsection{Luminosity}

The inferred effective temperature of S106 IR is consistent with a mid- to late-type O star \citep{vandenancker00}. 
However, the effective 
temperature of S106 IR determined by \citet{vandenancker00} using emission line intensities is uncertain due to opacity 
from the stellar/disk winds that are not included in their modeling, as well as any UV flux that is produced by accretion. 
Given this uncertainty, we cannot ascertain whether S106 IR is a massive binary or a single ZAMS star or pre-main sequence object.
For example, the dust luminosity is higher than the luminosity of a single ZAMS star with an effective temperature of 37,000 K 
($\sim 6.5 \times 10^4~\rm{L}_\odot$), according to the ZAMS published in \citet{yorke02} and \citet{zinnecker07}. The
effective temperature may be as high as 40,000 K, which could be produced by a single ZAMS star with luminosity 
$\sim 1.2 \times 10^5~\rm{L}_\odot$. Furthermore, S106 IR may be a pre-main sequence object on a Henyey
track, resulting in a larger radius and cooler temperature than those of a ZAMS star. Thus, we cannot establish the precise nature
of S106 IR in terms of binarity or evolutionary status.

\subsubsection{The circumstellar environment}

There are indications that S106 IR is a young stellar object. Its mass loss rate from winds is much higher than that of
a main sequence star \citep{hippelein81, felli84, drew93, hoare94, gibb07, lumsden12}. In addition, CO $\nu=2-0$ bandhead 
emission provides evidence for a disk on AU scales \citep{murakawa13}.
These indicators of youth allow us to place an upper limit on the age of S106 IR of $\sim 3 \times 10^5$ yr based on the timescale 
for photo-evaporation of disks around massive stars \citep{hollenbach00}.

The variability in dust temperatures in the equatorial region is consistent with a fragmented environment that is expected 
from the clumpy structure of the molecular gas observed by \citet{barsony89} and \citet{simon12} and dust fragments 
observed by \citet{richer93}. However, the anticorrelation between our values of $\tau_{UV}$ in the cool lanes and $D_{proj}$
might indicate that at least some of the extinction occurs in a small (AU scale) disk that may be also be clumpy.
The CO $\nu=2-0$ bandhead observations of \citet{murakawa13} imply the existence of a 
small disk. At the young age of
S106 IR, the dispersal of the disk occurs a few AU from the star and is dominated by viscous dispersal over timescales of
$10^4$ -- $10^5$ yr \citep{hollenbach00}. This suggests the possibility that the small disk may be a remnant disk or is dispersing or
that is being supplied with gas from the surrounding bar of cold dust and gas \citep{peters10}.

\subsection{ Dust heating }

It is likely that the star illuminates 
dust in the lobes through an opening in the equatorial geometry of the dust. This would explain the overall lower temperatures observed near 
S106 IR and relatively high temperatures
in the southern lobe. \citet{smith01} presented a color temperature map that exhibited a more uniform 
structure at a temperature of $\sim 135$ K, using a $\lambda^{-1}$ emissivity law. The SOFIA observations contain
sensitivity that extends farther into the lobes, where the temperature gradient is more apparent, which may 
explain the difference in temperature gradient between the two maps.

The agreement between the observed dust color temperatures and the equilibrium temperature computed by radiative
transfer in the southern lobe and PL 1, where stellar extinction is low, suggest that the dust heating is dominated by stellar radiation,
rather than by grain-electron collisions or trapped Ly-$\alpha$ heating. The following calculations support this assertion. 
We computed the radiative heating rate on a single grain assuming, conservatively, a moderate 
distance ($D = 0.15$ pc) from the star with some extinction of the UV field ($\tau_{UV} = 3$), and a grain aborption efficiency 
$Q_{\lambda}=1$ for $\lambda < 0.4~\mu$m and $Q_{\lambda} \propto \lambda^{-1.6}$ for $\lambda \ge 0.4~\mu$m \citep{draine84}.
The radiative heating rate, $\Gamma_{rad}$ is given by:
\begin{equation}
\Gamma_{rad} = \frac{R_*^2}{D^2} \pi a^2 \int F_{\lambda,*} e^{-\tau_\lambda} Q_\lambda d\lambda
\end{equation}
The heating rate $\Gamma_{coll}$ for grain-electron collisions is given by \citet{dwek87}: 
\begin{equation}
\Gamma_{coll} = \bigg( \frac{32}{\pi m_e} \bigg)^{1/2} \pi a^2 n_e (kT_e)^{3/2}
\end{equation}
where $m_e$ is the mass of an electron, $n_e$ is the number density of free electrons, and $T_e$ is the electron
temperature. We assumed a typical value for the electron temperatures 
in a gas-cooled nebula ($T_e \approx 10,000$) \citep{tielens05}
and used an estimated value for the electron number density ($n_e \approx 3 \times 10^4~\rm{cm}^{-3}$) from
Bally et al. (\citeyear{bally83}). We find the heating rate for grain-electron collisions is $\lesssim 15\%$ that of the 
radiative heating. Such a small component of the heating will have little effect on the grain equilibrium temperature.

We also expect a small contribution to the dust heating from Ly-$\alpha$ radiation due to the presence of warm gas 
(e.g. Schneider et al. \citeyear{schneider02}, \citeyear{schneider03}; Simon et al. \citeyear{simon12}). The rate of this heating 
$\Gamma_{\alpha}$ is given by \citet{tielens05}:
\begin{equation}
\Gamma_{\alpha} = \frac{n_e^2 \beta_B h \nu_\alpha}{n_d}
\end{equation}
where $\beta_B$ is the hydrogen recombination coefficient 
to all levels with $n \ge 2$, $\nu_\alpha$ is the frequency of a Ly-$\alpha$ photon, and $n_d$ is the number density of
dust particles. We assumed a high degree of ionization ($x \approx 1$), a charge neutral gas ($n_e = n_p$, where $n_p$ is the
number density of protons), and a standard gas-to-dust mass ratio of 100. For a grain with radius $0.1~\mu$m and
mass density $\rho = 3~\rm{g}~\rm{cm}^{-3}$, we find 
$\Gamma_{\alpha} \approx 3.2 \times 10^{-6}~\Gamma_{rad}$, where $\Gamma_{rad}$
is the heating rate on the grain from the star at a distance of 0.15 pc with $\tau_{UV} = 3$. Therefore, Ly-$\alpha$ radiation
will not appreciably affect the dust grain equilibrium temperatures.

\subsection{ Dust composition }

In the cool lane positions, the abundance ratio of amorphous carbon to amorphous silicate grains is variable, but brackets the values
found in the interstellar diffuse high galactic latitude fields of \citet{compiegne11} (83\% amorphous silicates, 17\% amorphous 
carbon). In
the regions with a stronger UV field, the relative abundances of the BGs, VSGs, and PAHs are similar to those found in the
ISM (88\% BGs, 6.4\% VSGs, 5.9\% PAHs) \citep{desert90} and in the M16 PDR locations \citep{flagey11}. The exception
is the relative VSG abundance in the southern lobe location SL 3 and the southwestern clump (PL 1) is higher than at locations 
closer to the star. The increase in VSGs is accompanied by a relatively high proportion of ionized PAHs when compared with the other locations. 
The increase in relative VSG abundance is similar to that in the ``reverse shell'' in M16 \citep{flagey11}; however the 
``reverse shell'' shows
an absence of PAHs. \citet{flagey11} provide wind-driven grain-grain collisions as one possible explanation for the enhancement of 
small grains in the M16 shell. Such a process may be at work at PL 1, although another explanation may be that there is less photo-evaporation
of VSGs as a consequence of the weaker radiation field at PL 1. Photo-evaporation of VSGs has been proposed to
explain the abundances of VSGs and PAHs in other regions such as Ced 201 \citep{cesarsky00, berne07}, NGC 2023 
North \citep{compiegne08}, $\rho$ Oph-SR3 and NGC 7023-NW \citep{rapacioli05}, and NGC 7023-E and $\rho$ Oph-filament
\citep{berne07}.

\subsection{Size of the nebular region}

H-$\alpha$ images show that the size of the H {\small II} region extends southwest from S106 IR to an ionization front near location PL 1 
(Bally et al. \citeyear{bally98}). 
Theory states that the Str\"{o}mgren radius of an H {\small II} region expands with time according to the equation \citep{wardthompson11}:
\begin{equation}
R(t) \simeq 5\left(\frac{\dot{N}_{HI}}{10^{50}~\rm{s}^{-1}}\right)^{1/7}\left(\frac{n_0}{10^3~\rm{cm}^{-3}}\right)^{-2/7}\left(\frac{t}{\rm{Myr}}\right)^{4/7}~\rm{pc}
\end{equation}
where $\dot{N}_{HI}$ is the number of hydrogen-ionizing photons per second emitted by the star, $n_0$ is the number density of gas molecules
in the surrounding molecular cloud, and $t$ is time.
If we assume $\dot{N}_{HI} = 4.1 \times 10^{48}~\rm{s}^{-1}$
\citep{sternberg03} and $n_0 \approx 1.4 \times 10^3~\rm{cm}^{-3}$ \citep{schneider02}, then
a Str\"{o}mgren radius of 0.6 pc (the distance from S106 IR to the ionization front assuming an inclination 
angle of $30^\circ$; Gehrz et al. \citeyear{gehrz82}) is reached in $\sim 5 \times 10^4$ yr. This timescale is consistent with the existence of a 
photo-evaporating disk, but we caution that it is only a rough estimate as $n_0$ may vary in the surrounding cloud.

Bipolar continuum emission extends beyond the ionization front. The size of the bipolar region depends on the 
accretion and outflow histories of S106 IR, and the clump at PL 1 may
be a remnant, higher density clump that is still eroding from the formation of the lobes through outflow. Alternatively, it may be a clump
of swept-up material at the ionization front. Either case indicates that there are complex dynamics in the bipolar region. Further study of the
dynamics of the region will require multi-wavelength, spectrally resolved observations.

\section{CONCLUSIONS}

We have used ground-, airborne, and space-based observations and performed radiation transfer modeling in order to study the 
warm dust in S106. We summarize the conclusions of this work as follows:
\begin{enumerate}
\item The total dust luminosity around S106 IR is $\gtrsim (9.02 \pm 1.01) \times 10^4~\rm{L}_\odot$. This luminosity is consistent
with the luminosity of a mid- to late-type O star; however, due to uncertainties, we cannot establish its precise nature in terms of binarity or
evolutionary status.
\item The dust temperature gradient ($\sim 75 - 107$ K) in the lobes is consistent with an equatorial geometry around S106 IR. The
dust is heated radiatively, with little contribution from grain-electron collisions and Ly-$\alpha$ radiation. 
\item Variable dust temperatures ($\sim 65 - 90$ K) near the equatorial plane {\bf ($D_{proj} \lesssim 0.1$ pc)} indicate
that the environment is fragmented ($\tau_{UV} = 2 - 40$).
\item The dust mass composition in the H {\small II} region is composed of BGs ($\sim 50 - 98\%$), with 
the remaining contributions from VSGs ($\sim 1 - 43\%$) and PAHs ($\lesssim 7\%$). The largest proportions of both VSGs 
and PAHs are located in the clump PL 1.
\end{enumerate}

\acknowledgments
\noindent We thank the SOFIA ground crew, flight crew, and Mission Operations for their successful execution of the SOFIA observations.
We also thank an anonymous referee for making suggestions that led to the improvement of this paper.
This work is based on observations made with the NASA/DLR Stratospheric Observatory for Infrared Astronomy (SOFIA). SOFIA science mission operations are conducted jointly by the Universities Space Research Association, Inc. (USRA), under NASA contract NAS2-97001, and the Deutsches SOFIA Institut (DSI) under DLR contract 50 OK 0901. Financial support for FORCAST was provided to Cornell by NASA through award 8500-98-014 issued by USRA.
This work is based on observations made with the {\it Spitzer Space Telescope}, which is operated by JPL/Caltech under NASA contract 1407. This work is based in part on observations made with Herschel, a European Space Agency Cornerstone Mission with significant participation by NASA. Support for this work was provided by NASA through an award issued by JPL/Caltech.
RDG acknowledges support from NASA and the United States Air Force.

{\facility {\it Facilities}: Spitzer, SOFIA, Herschel}

\newpage

\newpage

\newpage

\newpage

\newpage

\newpage

\newpage

\newpage

\newpage

\newpage

\newpage

\newpage
\begin{deluxetable}{ccc}
\tablecaption{\label{tab:totalflux} Total dust continuum flux densities in S106, including 350 $\mu$m data from \citet{simon12}. The values exclude emission from the point sources S106 IR and IRS 2 \citep{gehrz82}.}
\tablehead{\colhead{$\lambda$ ($\mu$m)} & \colhead{Aperture} & \colhead{$F_\nu$ (kJy)}}
\startdata
      12.492 & $1.^\prime01 \times 1.^\prime11$   &  $0.0649 \pm        0.0130$ \\
      19.7 &  $1.^\prime74 \times 2.^\prime30$ &  $2.53 \pm        0.25$ \\
      25.3 &  $1.^\prime74 \times 2.^\prime30$  &  $4.32 \pm        0.43$ \\
      31.5 &  $1.^\prime74 \times 2.^\prime30$  &  $5.66 \pm        0.57$ \\
      37.1 &  $1.^\prime74 \times 2.^\prime30$  &  $6.86 \pm        0.69$ \\
      70 &  $4.97^\prime \times 7.53^\prime$ elliptical  &  $19.1 \pm        1.9$ \\
      160 &  $4.97^\prime \times 7.53^\prime$ elliptical  &  $15.8 \pm        3.2$ \\
      350 &  $5^\prime.72 \times 4^\prime.05$ &  $0.630 \pm        0.126$ \\
\enddata
\end{deluxetable}

\newpage
\begin{deluxetable}{lrrrrrrrr}
\tabletypesize{\tiny}
\tablecaption{\label{tab:fluxes} Flux densities, in units of log Jy, at the locations indicated in Fig. \ref{fig:ctemp}. The table includes flux densities at 3.29 $\mu$m \citep{smith01}
and 350 $\mu$m \citep{simon12}.}
\tablehead{
\colhead{Location}
& \colhead{RA (2000)}
& \colhead{DEC}
& \colhead{3.29 $\mu$m}
& \colhead{3.6 $\mu$m}
& \colhead{4.5 $\mu$m}
& \colhead{5.8 $\mu$m}
& \colhead{8.0 $\mu$m}
& \colhead{11.3 $\mu$m}}
\startdata
CL 1 & 20:27:27.03  & +37:22:50.5 & $ -1.93 \pm  -3.00$ & $ -1.88 \pm  -5.12$ & $ -1.62 \pm  -5.28$ & $ -1.20 \pm  -4.51$ & $ -1.01 \pm  -4.09$ & $ -1.28 \pm  -1.97$ \\
CL 2 & 20:27:27.11  & +37:22:41.0 & $ -2.04 \pm  -3.00$ & $ -2.20 \pm  -5.12$ & $ -2.14 \pm  -5.28$ & $ -1.57 \pm  -4.51$ & $ -1.20 \pm  -4.09$ & $ -1.05 \pm  -1.97$ \\
CL 3 & 20:27:27.39  & +37:22:45.4 & $ -2.19 \pm  -3.00$ & $ -2.41 \pm  -5.12$ & $ -2.39 \pm  -5.28$ & $ -1.73 \pm  -4.51$ & $ -1.31 \pm  -4.09$ & $ -1.92 \pm  -1.97$ \\
CL 4 & 20:27:27.69  & +37:22:31.6 & $ -2.43 \pm  -3.00$ & $ -2.77 \pm  -5.12$ & $ -2.69 \pm  -5.28$ & $ -1.73 \pm  -4.51$ & $ -1.40 \pm  -4.09$ & $< -1.97$ \\
CL 5 & 20:27:27.67  & +37:22:53.1 & $ -2.90 \pm  -3.00$ & $ -2.70 \pm  -5.12$ & $ -2.53 \pm  -5.28$ & $ -1.90 \pm  -4.51$ & $ -1.47 \pm  -4.09$ & $ -1.33 \pm  -1.97$ \\
CL 6 & 20:27:28.25  & +37:22:54.1 & \nodata & $ -3.09 \pm  -5.12$ & $ -2.92 \pm  -5.28$ & $ -2.01 \pm  -4.51$ & $ -1.58 \pm  -4.09$ & $ -1.66 \pm  -1.97$ \\
CL 7 & 20:27:25.80  & +37:22:51.2 & $ -2.44 \pm  -3.00$ & $ -2.45 \pm  -5.12$ & $ -2.37 \pm  -5.28$ & $ -1.77 \pm  -4.51$ & $ -1.34 \pm  -4.09$ & $ -1.43 \pm  -1.97$ \\
WL 1  & 20:27:27.61  & +37:22:39.4 & $ -1.52 \pm  -3.00$ & $ -2.22 \pm  -5.60$ & $ -2.20 \pm  -5.76$ & $ -1.44 \pm  -4.99$ & $ -1.02 \pm  -4.57$ & $ -0.63 \pm  -1.97$ \\
WL 2  & 20:27:27.82  & +37:22:48.9 & $ -1.93 \pm  -3.00$ & $ -2.48 \pm  -5.60$ & $ -2.37 \pm  -5.76$ & $ -1.67 \pm  -4.99$ & $ -1.27 \pm  -4.57$ & $ -1.08 \pm  -1.97$ \\
WL 3  & 20:27:28.33  & +37:22:48.1 & $ -2.45 \pm  -3.00$ & $ -2.72 \pm  -5.60$ & $ -2.65 \pm  -5.76$ & $ -1.73 \pm  -4.99$ & $ -1.34 \pm  -4.57$ & $ -1.47 \pm  -1.97$ \\
IRS 1  & 20:27:25.24  & +37:22:32.2 & $ -2.39 \pm  -3.00$ & $ -2.74 \pm  -5.60$ & $ -2.69 \pm  -5.76$ & $ -1.99 \pm  -4.99$ & $ -1.54 \pm  -4.57$ & $ -1.09 \pm  -1.97$ \\
IRS 3  & 20:27:25.66  & +37:22:42.6 & $ -1.41 \pm  -3.00$ & $ -1.96 \pm  -5.60$ & $ -1.84 \pm  -5.76$ & $ -1.19 \pm  -4.99$ & $ -0.75 \pm  -4.57$ & $ -0.29 \pm  -1.97$ \\
IRS 5  & 20:27:26.80  & +37:22:57.3 & $ -1.85 \pm  -3.00$ & $ -2.11 \pm  -5.60$ & $ -2.00 \pm  -5.76$ & $ -1.45 \pm  -4.99$ & $ -0.95 \pm  -4.57$ & $ -0.51 \pm  -1.97$ \\
IRS 6  & 20:27:27.19  & +37:22:29.0 & $ -1.42 \pm  -3.00$ & $ -2.00 \pm  -5.60$ & $ -1.96 \pm  -5.76$ & $ -1.24 \pm  -4.99$ & $ -0.80 \pm  -4.57$ & $ -0.27 \pm  -1.97$ \\
IRS 7  & 20:27:27.40  & +37:22:39.3 & $ -1.66 \pm  -3.00$ & $ -2.22 \pm  -5.60$ & $ -2.13 \pm  -5.76$ & $ -1.44 \pm  -4.99$ & $ -1.01 \pm  -4.57$ & $ -0.54 \pm  -1.97$ \\
IRS 8  & 20:27:27.45  & +37:23:00.9 & $ -1.82 \pm  -3.00$ & $ -2.25 \pm  -5.60$ & $ -2.18 \pm  -5.76$ & $ -1.48 \pm  -4.99$ & $ -1.03 \pm  -4.57$ & $ -0.86 \pm  -1.97$ \\
SL 1  & 20:27:25.96  & +37:22:28.8 & $ -2.15 \pm  -3.00$ & $ -2.55 \pm  -5.60$ & $ -2.56 \pm  -5.76$ & $ -1.79 \pm  -4.99$ & $ -1.37 \pm  -4.57$ & $ -1.84 \pm  -1.97$ \\
SL 2  & 20:27:26.46  & +37:22:35.8 & $ -1.83 \pm  -3.00$ & $ -2.21 \pm  -5.60$ & $ -2.17 \pm  -5.76$ & $ -1.55 \pm  -4.99$ & $ -1.12 \pm  -4.57$ & $ -0.82 \pm  -1.97$ \\
SL 3  & 20:27:25.25  & +37:22:21.0 & $ -2.52 \pm  -3.00$ & $ -3.06 \pm  -5.60$ & $ -2.93 \pm  -5.76$ & $ -2.33 \pm  -4.99$ & $ -1.90 \pm  -4.57$ & $ -1.78 \pm  -1.97$ \\
PL 1  & 20:27:23.84  & +37:21:37.8 & $ -2.11 \pm  -3.00$ & $ -2.58 \pm  -5.60$ & $ -2.62 \pm  -5.76$ & $ -1.76 \pm  -4.99$ & $ -1.35 \pm  -4.57$ & \nodata \\
\hline \hline
Location  & 12.492 $\mu$m & 19.7 $\mu$m & 25.3 $\mu$m & 31.5 $\mu$m & 37.1 $\mu$m & 70 $\mu$m & 160 $\mu$m & 350 $\mu$m \\
\hline
CL 1 & $ -1.53 \pm  -2.67$ & $ -0.19 \pm  -1.18$ & $  0.27 \pm  -0.73$ & $  0.44 \pm  -0.56$ & $  0.62 \pm  -0.38$ & $  0.79 \pm  -0.21$ & $  0.30 \pm  -0.40$ & $ -0.61 \pm  -2.19$ \\
CL 2 & $ -1.40 \pm  -2.67$ & $ -0.11 \pm  -1.11$ & $  0.37 \pm  -0.63$ & $  0.62 \pm  -0.38$ & $  0.74 \pm  -0.26$ & $  0.77 \pm  -0.23$ & $  0.24 \pm  -0.46$ & $ -0.86 \pm  -2.19$ \\
CL 3 & $ -1.74 \pm  -2.67$ & $ -0.39 \pm  -1.38$ & $  0.03 \pm  -0.97$ & $  0.33 \pm  -0.67$ & $  0.49 \pm  -0.51$ & $  0.77 \pm  -0.23$ & $  0.31 \pm  -0.39$ & $ -0.70 \pm  -2.19$ \\
CL 4 & $< -2.67$ & $ -0.77 \pm  -1.73$ & $ -0.29 \pm  -1.28$ & $  0.04 \pm  -0.96$ & $ -0.00 \pm  -1.00$ & $  0.60 \pm  -0.40$ & $  0.14 \pm  -0.55$ & $ -0.88 \pm  -2.19$ \\
CL 5 & \nodata & $ -0.36 \pm  -1.36$ & $  0.08 \pm  -0.91$ & $  0.36 \pm  -0.64$ & $  0.46 \pm  -0.54$ & $  0.79 \pm  -0.21$ & $  0.34 \pm  -0.36$ & $ -0.51 \pm  -2.19$ \\
CL 6 & $ -2.01 \pm  -2.67$ & $ -0.63 \pm  -1.61$ & $ -0.11 \pm  -1.11$ & $  0.16 \pm  -0.84$ & $  0.16 \pm  -0.84$ & $  0.62 \pm  -0.38$ & $  0.29 \pm  -0.41$ & $ -0.49 \pm  -2.19$ \\
CL 7 & $ -1.91 \pm  -2.67$ & $ -0.38 \pm  -1.37$ & $ -0.03 \pm  -1.03$ & $  0.33 \pm  -0.67$ & $  0.53 \pm  -0.47$ & $  0.77 \pm  -0.23$ & $  0.31 \pm  -0.39$ & $ -0.38 \pm  -2.19$ \\
WL 1 & $ -0.98 \pm  -2.67$ & $  0.25 \pm  -0.75$ & $  0.59 \pm  -0.41$ & $  0.76 \pm  -0.24$ & $  0.65 \pm  -0.35$ & $  0.72 \pm  -0.28$ & $  0.23 \pm  -0.47$ & $ -0.74 \pm  -2.19$ \\
WL 2 & $ -1.27 \pm  -2.67$ & $ -0.12 \pm  -1.12$ & $  0.27 \pm  -0.73$ & $  0.44 \pm  -0.56$ & $  0.45 \pm  -0.55$ & $  0.72 \pm  -0.28$ & $  0.32 \pm  -0.38$ & $ -0.63 \pm  -2.19$ \\
WL 3 & $ -1.74 \pm  -2.67$ & $ -0.36 \pm  -1.35$ & $  0.03 \pm  -0.97$ & $  0.14 \pm  -0.86$ & $  0.14 \pm  -0.86$ & $  0.50 \pm  -0.50$ & $  0.20 \pm  -0.49$ & $ -0.75 \pm  -2.19$ \\
IRS 1 & $ -1.35 \pm  -2.67$ & $ -0.42 \pm  -1.42$ & $ -0.29 \pm  -1.29$ & $ -0.21 \pm  -1.21$ & $ -0.12 \pm  -1.12$ & $ -0.01 \pm  -1.01$ & $ -0.36 \pm  -1.06$ & $< -2.19$ \\
IRS 3 & $ -0.57 \pm  -2.67$ & $  0.58 \pm  -0.42$ & $  0.82 \pm  -0.18$ & $  0.91 \pm  -0.09$ & $  0.90 \pm  -0.10$ & $  0.72 \pm  -0.28$ & $  0.24 \pm  -0.46$ & $ -1.02 \pm  -2.19$ \\
IRS 5 & $ -0.85 \pm  -2.67$ & $  0.36 \pm  -0.64$ & $  0.59 \pm  -0.41$ & $  0.64 \pm  -0.36$ & $  0.74 \pm  -0.26$ & $  0.63 \pm  -0.37$ & $  0.15 \pm  -0.55$ & $ -0.77 \pm  -2.19$ \\
IRS 6 & $ -0.69 \pm  -2.67$ & $  0.44 \pm  -0.56$ & $  0.68 \pm  -0.32$ & $  0.71 \pm  -0.29$ & $  0.71 \pm  -0.29$ & $  0.47 \pm  -0.53$ & $ -0.05 \pm  -0.75$ & $ -1.38 \pm  -2.19$ \\
IRS 7 & $ -0.84 \pm  -2.67$ & $  0.23 \pm  -0.77$ & $  0.65 \pm  -0.35$ & $  0.78 \pm  -0.22$ & $  0.76 \pm  -0.24$ & $  0.75 \pm  -0.25$ & $  0.24 \pm  -0.46$ & $ -0.78 \pm  -2.19$ \\
IRS 8 & $ -0.93 \pm  -2.67$ & $  0.41 \pm  -0.59$ & $  0.63 \pm  -0.37$ & $  0.66 \pm  -0.34$ & $  0.71 \pm  -0.29$ & $  0.62 \pm  -0.38$ & $  0.09 \pm  -0.61$ & $ -0.89 \pm  -2.19$ \\
SL 1 & $ -1.65 \pm  -2.67$ & $ -0.47 \pm  -1.47$ & $ -0.25 \pm  -1.25$ & $ -0.08 \pm  -1.08$ & $ -0.05 \pm  -1.05$ & $  0.12 \pm  -0.88$ & $ -0.29 \pm  -0.99$ & $ -1.94 \pm  -2.19$ \\
SL 2 & $ -1.15 \pm  -2.67$ & $  0.03 \pm  -0.97$ & $  0.34 \pm  -0.66$ & $  0.50 \pm  -0.50$ & $  0.55 \pm  -0.45$ & $  0.49 \pm  -0.51$ & $  0.01 \pm  -0.69$ & $ -1.18 \pm  -2.19$ \\
SL 3 & $ -1.61 \pm  -2.67$ & $ -0.72 \pm  -1.71$ & $ -0.67 \pm  -1.65$ & $ -0.80 \pm  -1.77$ & $ -0.67 \pm  -1.64$ & $ -0.41 \pm  -1.41$ & $ -0.69 \pm  -1.39$ & \nodata \\
PL 1 & \nodata & $ -0.86 \pm  -1.84$ & $ -0.57 \pm  -1.56$ & $ -0.34 \pm  -1.34$ & $ -0.28 \pm  -1.28$ & $ -0.12 \pm  -1.12$ & $ -0.59 \pm  -1.29$ & $ -2.08 \pm  -2.19$ \\
\enddata
\end{deluxetable}

\newpage
\begin{deluxetable}{lccccccccc}
\tabletypesize{\small}
\tablecaption{\label{tab:models1}Positions, model parameters, and model
results for cool lanes CL 1--7. $D_{proj}$ is the projected distance between the specified location
and S106 IR, scaled to a distance of 1.4 kpc \citep{schneider07}. The aperture used 
to derive the observed SEDs (Fig. \ref{fig:models1}) for the cool lane positions was 
$0.863^{\prime\prime} \times 0.863^{\prime\prime}$. 
Also listed are values
of $\tau_{UV}$, $G_0$ in units of the Habing field, power law size distribution indices for amorphous carbon and amorphous
silicates ($\alpha_C$ and $\alpha_{Si}$, respectively) and dust mass fractions for neutral PAHs, ionized PAHs,
amorphous carbon and amorphous silicates ($f_{PAH0}$, $f_{PAH1}$, $f_C$ and $f_{Si}$, respectively), normalized to 1.}
\tablehead{
\colhead{Position}
& \colhead{$D_{proj}$ (pc)}
& \colhead{$\tau_{UV}$}
& \colhead{$G_0$\tablenotemark{a} (Habing)}
& \colhead{$\alpha_C$}
& \colhead{$\alpha_{Si}$}
& \colhead{$f_{PAH0}$}
& \colhead{$f_{PAH1}$}
& \colhead{$f_C$}
& \colhead{$f_{Si}$}
}
\startdata
CL 1  & 0.021 & 40  & 2270 & -3.10  & -3.50  & \nodata & \nodata & 0.300 & 0.700 \\
CL 2  & 0.057 & 17  & 1030 & -3.20  & -3.50  & 0.002 & 0.001 & 0.488 & 0.509 \\
CL 3  & 0.052 & 20  & 997 & -3.25  & -3.30  & \nodata & \nodata & 0.150 & 0.850 \\
CL 4  & 0.14 & 8  & 463 & -3.45  & -3.40 & \nodata & 0.001 & 0.062 & 0.936 \\
CL 5  & 0.076 & 16  & 650 & -3.40  & -3.50  & \nodata & 0.003 & 0.249 & 0.748 \\
CL 6  & 0.12 & 13  & 325 & -3.22  & -3.50  & \nodata & 0.002 & 0.499 & 0.499 \\
CL 7  & 0.083 & 17  & 494 & -3.10  & -3.50  & $<$ 0.001 & 0.002 & 0.499 & 0.499 \\
\enddata
\tablenotetext{a}{Computed using $\tau_{UV}$ and $D_{proj}$.}
\vspace{7.0cm}
\end{deluxetable}
\vspace{7.0cm}

\newpage
\begin{deluxetable}{lccccccccc}
\tabletypesize{\small}
\tablecaption{\label{tab:models2}Positions, model parameters, and model
results for warm lane positions WL 1--3, extended IRS sources from \citet{gehrz82}, southern lobe
positions SL 1--3, and southwestern clump position PL 1. $D_{proj}$ is the projected distance between the specified location
and S106 IR, scaled to a distance of 1.4 kpc \citep{schneider07}. The aperture used 
to derive the observed SEDs (Fig. \ref{fig:models1} and \ref{fig:models2}) for these positions was 
$\sim 2.6^{\prime\prime} \times 2.6^{\prime\prime}$. 
Also listed are values of $\tau_{UV}$, $G_0$ in units of the Habing field,power law size distribution indices for VSGs and 
BGs ($\alpha_{VSG}$ and $\alpha_{BG}$, respectively) 
and dust mass fractions for neutral PAHs, ionized PAHs, VSGs, and BGs ($f_{PAH0}$,  $f_{PAH1}$, $f_{VSG}$, and
$f_{BG}$, respectively), normalized to 1.}
\tablehead{
\colhead{Position}
& \colhead{$D_{proj}$ (pc)}
& \colhead{$\tau_{UV}$}
& \colhead{$G_0$\tablenotemark{a} (Habing)}
& \colhead{$\alpha_{VSG}$}
& \colhead{$\alpha_{BG}$}
& \colhead{$f_{PAH0}$}
& \colhead{$f_{PAH1}$}
& \colhead{$f_{VSG}$}
& \colhead{$f_{BG}$}
}
\startdata
WL 1  & 0.091 & 4.0  & 2740 & -2.0  & -2.9  & 0.003 & 0.014 & 0.004 & 0.979 \\
WL 2  & 0.081 & 6.0  & 1850 & -2.6  & -2.9  & 0.003 & 0.021 & 0.005 & 0.972 \\
WL 3  & 0.12 & 2.0  & 4040 & -2.6  & -2.9  & $<$ 0.001 & 0.014 & 0.005 & 0.981 \\
IRS 1  & 0.17 & 1.0  & 3980 & -2.6  & -2.9  & 0.002 & 0.027 & 0.061 & 0.910 \\
IRS 3  & 0.10 & 2.0  & 5840 &-2.6  & -2.9  & 0.002 & 0.016 & 0.027 & 0.955 \\
IRS 5  & 0.058 & 5.0  & 4700 & -2.3  & -2.9  & 0.002 & 0.024 & 0.049 & 0.925 \\
IRS 6  & 0.14 & 0.3  & 9670 & -2.6  & -2.0  & $<$ 0.001 & 0.008 & 0.003 & 0.989 \\
IRS 7  & 0.079 & 3.0  & 5680 & -2.6  & -2.0  & 0.002 & 0.013 & 0.005 & 0.981 \\
IRS 8  & 0.097 & 2.0  & 6480 & -2.6  & -2.9  & 0.001 & 0.013 & 0.009 & 0.977 \\
SL 1  & 0.15 & 1.5  & 3620 & -2.6  & -2.9  & 0.002 & 0.033 & 0.011 & 0.954 \\
SL 2  & 0.092 & 3.3  & 3640 & -2.6  & -2.9  & 0.002 & 0.021 & 0.007 & 0.970 \\
SL 3  & 0.23 & 0.0  & 4380  & -3.1 & -2.9  & 0.001 & 0.013 & 0.106 & 0.880 \\
PL 1  & 0.54 & 0.0  & 781  & -2.6 & -2.9  & 0.002 & 0.067 & 0.278 & 0.668 \\
\enddata
\tablenotetext{a}{Computed using $\tau_{UV}$ and $D_{proj}$.}
\end{deluxetable}


\newpage

\begin{thebibliography}{ }

\bibitem[Abergel et al.(2011)]{abergel11}
Abergel, A., Ade, P. A. R., Aghanim, N. et al. 2011, \aap, 536, A25

\bibitem[Adams et al.(2012a)]{adams12a}
Adams, J. D., Herter, T. L., Gull, G. E. et al. 2012a, \procspie, 8446, 16

\bibitem[Adams et al.(2012b)]{adams12b}
Adams, J. D., Herter, T. L., Osorio, M. et al. 2012b, \apjl, 749L, 24

\bibitem[Allen \& Penston(1975)]{allen75}
Allen, D. A. \& Penston, M. V. 1975, \mnras, 172, 245

\bibitem[Bally \& Scoville(1982)]{bally82}
Bally, J. \& Scoville, N. Z. 1982, \apj, 255, 497

\bibitem[Bally et al. (1983)]{bally83}
Bally, J., Snell, R. L., \& Predmore, R. 1983, \apj, 272, 154

\bibitem[Bally et al.(1998)]{bally98}
Bally, J., Yu, K. C., Rayner, J, \& Zinnecker, H. 1998, \aj, 116, 1868

\bibitem[Barsony et al.(1989)]{barsony89}
Barsony, M., Scoville, N. Z., Bally, J., \& Claussen, M. J. 1989, \apj, 343, 212

\bibitem[Bern\'{e} et al.(2007)]{berne07}
Bern\'{e}, O., Joblin, C., Deville, Y. et al. 2007, \aap, 469, 575

\bibitem[Bieging(1984)]{bieging84}
Bieging, J. H. 1984, \apj, 286, 591

\bibitem[Castelli \& Kurucz(2004)]{castelli04}
Castelli, F. \& Kurucz, R. L. 2004, arXiv:0405087

\bibitem[Castor, McCray, \& Weaver(1975)]{castor75}
Castor, J., McCray, R., \& Weaver, R. 1975, \apjl, 200, L107

\bibitem[Cesarsky et al.(2000)]{cesarsky00}
Cesarsky, D., Lequeux, J., Ryter, C., \& G\'{e}rin, M. 2000, \aap, 354, L87

\bibitem[Compi\`{e}gne et al.(2008)]{compiegne08}
Compi\`{e}gne, M., Abergel, A., Verstraete, L., \& Habart, E. 2008, \aap, 491, 797

\bibitem[Compi\`{e}gne et al.(2010)]{compiegne10}
Compi\`{e}gne, M., Flagey, N., Noriega-Crespo, A. et al. 2010, \apj, 724, 44

\bibitem[Compi\`{e}gne et al.(2011)]{compiegne11}
Compi\`{e}gne, M., Verstraete, L., Jones, A. et al. 2011, \aap, 525, 103

\bibitem[De Buizer et al.(2012)]{debuizer12}
De Buizer, J. M., Morris, M. R., Beckline, E. E. et al. 2012, \apj, 749L, 23

\bibitem[D\'{e}sert, Boulanger, \& Puget(1990)]{desert90}
D\'{e}sert, F.-X., Boulanger, F., \& Puget, J. L. 1990, \aap, 237, 215

\bibitem[Draine \& Lee(1984)]{draine84}
Draine, B. T. \& Lee, H. M. 1984, \apj, 285, 89

\bibitem[Draine \& Li(2007)]{draine07}
Draine, B. T. \& Li, A. 2007, \apj, 657, 810

\bibitem[Drew et al.(1993)]{drew93}
Drew, J. E., Bunn, J. C., \& Hoare, M. G. 1993, \mnras, 265, 12

\bibitem[Dwek(1987)]{dwek87}
Dwek, E. 1987, \apj, 322, 812


\bibitem[Eiroa, Els\"{a}sser, \& Lahulla(1979)]{eiroa79}
Eiroa, C., Els\"{a}sser, H., \& Lahulla, J. F. 1979, \aap, 74, 89

\bibitem[Fazio et al.(2004)]{fazio04}
Fazio, G. et al. 2004, \apjs, 154, 10

\bibitem[Felli et al.(1984)]{felli84}
Felli, M., Staude, H. J., Reddmann, T. et al. 1984, \aap, 135, 261

\bibitem[Flagey et al.(2011)]{flagey11}
Flagey, N., Boulanger, F., Noriega-Crespo, A. et al. 2011, \aap, 531, A51

\bibitem[Gehrz et al.(1982)]{gehrz82}
Gehrz, R. D., Grasdalen, G. L., Castelaz, M. et al. 1982, \apj, 254, 550

\bibitem[Gibb \& Hoare(2007)]{gibb07}
Gibb, A. G. \& Hoare, M. G. 2007, \mnras, 380, 246

\bibitem[Griffin et al.(2010)]{griffin10}
Griffin, M. J., Abergel, A., Abreu, A. et al. 2010, \aap, 518L, 3

\bibitem[Herter et al.(1982)]{herter82}
Herter, T., Helfer, H. L., Pipher, J. L. et al. 1982, \apj, 262, 153

\bibitem[Herter et al.(2012)]{herter12}
Herter, T. L., Adams, J. D., De Buizer, J. M. et al. 2012, \apjl, 749L, 18

\bibitem[Herter et al.(2013)]{herter13}
Herter, T. L., Vacca, W. D., Adams, J. D. et al. 2013, \pasp, 125, 1393

\bibitem[Hippelein \& M\"{u}nch(1981)]{hippelein81}
Hippelein, H. \& M\"{u}nch, G. 1981, \aap, 99, 248

\bibitem[Hoare et al.(1994)]{hoare94}
Hoare, M. G., Drew, J. E., Muxlow, T. B., \& Davis, R. J. 1994, \apjl, 421, L51

\bibitem[Hodapp \& Rayner(1991)]{hodapp91}
Hodapp, K.-W. \& Rayner, J. 1991, \aj, 102, 1108

\bibitem[Hollenbach, Yorke, \& Johnstone(2000)]{hollenbach00}
Hollenbach, D. J., Yorke, H. W., \& Johnstone, D. 2000, in Protostars and Planets IV, ed. Mannings I., Boss A. P., \& Russell S. S. (Tucson, AZ: Univ. Ariz. Press), 401

\bibitem[Hora et al.(2004)]{hor04} 
Hora, J. et al. 2004, in Mather J., ed., Proc. of the SPIE: Optical, Infrared, and Millimeter Space Telescopes, Vol. 5487, p. 77

\bibitem[Hora et al.(2009)]{hora09}
Hora, J. L., Bontemps, S., Megeath, S. T. et al. 2009, \baas, 213, 356.01

\bibitem[Israel \& Felli(1978)]{israel78}
Israel, F. J. \& Felli, M. 1978, \aap, 63, 325

\bibitem[Kraemer et al.(2010)]{kraemer10}
Kraemer, K. E., Hora, J. L., Adams, J. D. et al. 2010, \baas, 215, 414.01

\bibitem[Landini et al.(1984)]{landini84}
Landini, M., Natta, A., Oliva, E. et al. 1984, \aap, 134, 284

\bibitem[Lucas et al.(1978)]{lucas78}
Lucas, A. M., Le Squeren, A. M., Kazes, I. et al. 1978, \aap, 66, 155

\bibitem[Lucy(1974)]{lucy74}
Lucy, L. B. 1974, \aj, 79, 745

\bibitem[Lumsden et al.(2012)]{lumsden12}
Lumsden, S. L., Wheelwright, H. E., Hoare, M. G. et al. 2012, \mnras, 424, 1088

\bibitem[Makovoz \& Khan(2005)]{mah05} 
Makovoz, D. \& Khan, I. 2005, in ASP Conf. Ser. 132, Astronomical Data Analysis Software and Systems VI, ed. P.L. Shopbell, M. C. Britton, \& R. Ebert (San Francisco, CA: ASP), 81

\bibitem[Martins, Schaerer, \& Hillier(2005)]{martins05}
Martins, F., Schaerer, D., \& Hillier, D. J. 2005, \aap, 436, 1039

\bibitem[Mason et al.(1998)]{mason98}
Mason, B. D., Gies, D. R., Hartkopf, W. I. 1998, \aj, 115, 821

\bibitem[Mezger et al.(1987)]{mezger87}
Mezger, P. G., Chini, R., Kreysa, E., \& Wink, J. 1987, \aap, 182, 127

\bibitem[Murakawa et al.(2013)]{murakawa13}
Murakawa, K., Lumsden, S. L., Oudmaijer, R. D. et al. 2013, \mnras, 436, 511

\bibitem[Oasa et al.(2006)]{oasa06}
Oasa, Y., Tamura, M., Nakajima, Y et al. 2006, \aj, 131, 1608

\bibitem[Panagia(1973)]{panagia73}
Panagia, N. 1973, \aj, 78, 929

\bibitem[Peters et al.(2010)]{peters10}
Peters, T., Banerjee, R., Klessen, R. S. et al. 2010, \apj, 711, 1017

\bibitem[Pipher et al.(1978)]{pipher78}
Pipher, J. L., Sharpless, S., Savedoff, M. P. et al. 1978, \aap, 59, 215

\bibitem[Pipher et al.(2004)]{pip04} 
Pipher, J. L. et al. 2004, in Proc. SPIE 5487, Optical, Infrared, and Millimeter Space Telescopes, ed. J. Mather J., 234

\bibitem[Poglitsch et al.(2010)]{poglitsch10}
Poglitsch, A., Waelkens, C., Geis, N. et al. 2010, \aap, 518L, 2

\bibitem[Price et al.(2001)]{price01}
Price, S. D., Egan, M. P., Carey, S. J. et al. 2001, \aj, 121, 2819

\bibitem[Rapacioli, Joblin, \& Boissel(2005)]{rapacioli05}
Rapacioli, M., Joblin, C., \& Boissel, P. 2005, \aap, 429, 193

\bibitem[Ressler et al.(1994)]{ressler94}
Ressler, M. E., Werner, M. W., van Cleve, J., \& Chou, H. A. 1994, in Infrared Astronomy with Arrays, the Next Generation 190, ed. I. S. McLean, Astrophysics and Space Sciences Library, 429

\bibitem[Richardson(1972)]{richardson72}
Richardson, W. H. 1972, JOSA, 62, 55

\bibitem[Richer et al.(1993)]{richer93}
Richer, J. S., Padman, R., Ward-Thompson, D. et al. 1993, \mnras, 262, 839

\bibitem[Roussel(2013)]{roussel13}
Roussel, H. 2013, \pasp, 125, 1126

\bibitem[Ryter, Puget, \& P\'{e}rault(1987)]{ryter87}
Ryter, C., Puget, J. L., \& P\'{e}rault, M. 1987, \aap, 186, 312

\bibitem[Schaerer et al.(1996)]{schaerer96}
Schaerer, D., de Koter, A., Schmutz, W., \& Maeder, A. 1996, \aap, 310, 837

\bibitem[Schuster et al.(2006)]{sch06} 
Schuster, M. T., Marengo, M., \& Patten, B. M. 2006, Proc. SPIE, 6270, 65

\bibitem[Schneider et al.(2002)]{schneider02}
Schneider, N., Simon, R., Kramer, C. et al. 2002, \aap, 384, 225

\bibitem[Schneider et al.(2003)]{schneider03}
Schneider, N., Simon, R., Kramer, C. et al. 2003, \aap, 406, 915

\bibitem[Schneider et al.(2007)]{schneider07}
Schneider, N., Simon, R., Bontemps, S. et al. 2007, \aap, 474, 873

\bibitem[Sharpless(1959)]{sharpless59}
Sharpless, S. 1959, A\&AS, 4, 257

\bibitem[Sibille et al.(1975)]{sibille75}
Sibille, F., Bergeat, J., Lunel, M., \& Kandel, R. 1975, \aap, 40, 441

\bibitem[Simon et al.(2012)]{simon12}
Simon, R., Schneider, N., Stutzki, J. et al. 2012, \aap, 542, L12

\bibitem[Smith et al.(2001)]{smith01}
Smith, N., Jones, T. J., Gehrz, R. D. et al. 2001, \aj, 121, 984

\bibitem[Stasi\'{n}ska \& Schaerer(1997)]{stasinska97}
Stasi\'{n}ska, G. \& Schaerer, D. 1997, \aap, 322, 615

\bibitem[Staude et al.(1982)]{staude82}
Staude, H. J., Lenzen, R., Dyck, H. M., \& Schmidt, G. D. 1982, \apj, 255, 95

\bibitem[Sternberg, Hoffmann, \& Pauldrach(2003)]{sternberg03}
Sternberg, A., Hoffmann, T. L., \& Pauldrach, A. W. A. 2003, \apj, 599, 1333

\bibitem[Tielens(2005)]{tielens05}
Tielens, A. G. G. M. 2005, The Physics and Chemistry of the Interstellar Medium, (Cambridge: Cambridge University Press)

\bibitem[Tokunaga \& Thompson(1979)]{tokunaga79}
Tokunaga, A. \& Thompson, R. 1979, \apj, 231, 736


\bibitem[van den Ancker et al.(2000)]{vandenancker00}
van den Ancker, M. E., Tielens, A. G. G. M., \& Wesselius, P. R. 2000, \aap, 358, 1035

\bibitem[Ward-Thompson \& Whitworth(2011)]{wardthompson11}
Ward-Thompson, D. \& Whitworth, A. P. 2011, An Introduction to Star Formation, (New York: Cambridge University Press)

\bibitem[Werner et al.(2004)]{werner04}
Werner, M. et al. 2004, \apjs, 154, 1

\bibitem[Yorke \& Sonnhalter(2002)]{yorke02}
Yorke, H. W. \& Sonnhalter, C. 2012, \apj, 569, 846

\bibitem[Young et al.(2012)]{young12}
Young, E. T., Becklin, E. E., De Buizer, J. M. et al. 2012, \apjl, 749L, 17

\bibitem[Zinnecker \& Yorke(2007)]{zinnecker07}
Zinnecker, H. \& Yorke, H. W. 2007, \araa, 45, 481

\end{thebibliography}
\end{document}